\documentclass[3p,nonatbib,10pt]{elsarticle}
\usepackage[utf8]{inputenc}
\usepackage[T1]{fontenc}
\usepackage[hypertexnames=false]{hyperref}
\usepackage{bm}

\usepackage{amsmath}
\usepackage{mathrsfs}
\usepackage{amsfonts}
\usepackage{amssymb}
\usepackage{mathtools}
\usepackage{array}
\usepackage{fancybox}
\usepackage{bm}
\usepackage{upgreek}
\usepackage[dvipsnames]{xcolor}
\usepackage{subcaption}
\usepackage{booktabs}
\usepackage{todonotes}
\usepackage{lineno}
\usepackage{comment}
\usepackage{makecell}
\usepackage{setspace} 
\usepackage{empheq}

\usepackage{pdfpages}
\usepackage{lipsum}
\usepackage{xspace}
\usepackage{xcolor}

\usepackage{biblatex}
\usepackage{multirow}
\usepackage{multicol}

\usepackage{nicematrix}

\newcolumntype{L}{>{$}l<{$}} 
\newcolumntype{R}{>{$}r<{$}} 
\newcolumntype{C}{>{$}c<{$}} 

\usepackage[font=small]{caption}

\newcommand*{\barbI}[1]{#1} 
\newcommand*{\barbII}[1]{#1} 
\newcommand*{\barbIII}[1]{#1}

\newcommand{\blat}[1]{\ensuremath{\mathbf{#1}}}
\newcommand{\bgre}[1]{\ensuremath{\bm{#1}}}

\newcommand{\boldc}{\barbI{\blat{c}}}
\newcommand{\boldC}{\barbII{\blat{C}}}

\newcommand{\bolde}{\barbI{\blat{e}}}
\newcommand{\boldf}{\barbI{\blat{f}}}
\newcommand{\ft}{\ensuremath{f_{\mathrm{t}}}}
\newcommand{\feff}{\ensuremath{f_{\mathrm{eq}}}}
\newcommand{\teff}{\ensuremath{t_{\mathrm{eq}}}}
\newcommand{\eeff}{\ensuremath{e_{\mathrm{eq}}}}

\newcommand{\Gt}{\ensuremath{G_{\mathrm{t}}}}

\newcommand{\boldI}{\barbII{\blat{I}}}

\newcommand{\boldn}{\barbI{\blat{n}}}

\newcommand{\boldt}{\barbI{\blat{t}}}

\newcommand{\boldu}{\barbI{\blat{u}}}

\newcommand{\boldx}{\barbI{\blat{x}}}

\newcommand{\boldtheta}{\barbI{\bgre{\uptheta}}}

\newcommand{\boldvarepsilon}{\barbII{\bgre{\upvarepsilon}}}

\newcommand{\boldxi}{\barbI{\bgre{\upxi}}}
\newcommand{\boldlambda}{\barbII{\bgre{\uplambda}}}
\newcommand{\boldpsi}{\barbII{\bgre{\uppsi}}}

\newcommand{\boldsigma}{\barbII{\bgre{\upsigma}}}

\newcommand{\levicivita}{\mathcal{E}}
\newcommand{\boldlevicivita}{\barbIII{\bm{\mathcal{E}}}}

\newcommand{\effE}{\ensuremath{E_{\mathrm{eff}}}}
\newcommand{\effnu}{\ensuremath{\nu_{\mathrm{eff}}}}
\newcommand{\lcorr}{\ensuremath{\ell_{\mathrm{c}}}}
\newcommand{\lRVE}{\ensuremath{l_{\mathrm{RVE}}}}

\newcommand{\matS}{{\sf S}\xspace}
\newcommand{\matA}{{\sf A}\xspace}
\newcommand{\matH}{{\sf H}\xspace}
\newcommand{\matV}{{\sf V}\xspace}
\newcommand{\matVRD}{{\sf V-RD}\xspace}
\newcommand{\matHR}{{\sf H-R}\xspace}
\newcommand{\matHRD}{{\sf H-RD}\xspace}
\newcommand{\matHRF}{{\sf H-RF}\xspace}

\let\svthefootnote\thefootnote
\newcommand\freefootnote[1]{%
  \let\thefootnote\relax%
  \footnotetext{#1}%
  \let\thefootnote\svthefootnote%
}

\journal{Engineering Fracture Mechanics}

\begin{document}
\begin{frontmatter}

\title{Simulating Heterogeneity within Elastic and Inelastic\\ Discrete Mechanical Models}

\auth[1]{Jan Raisinger}
\address[1]{Institute of Structural Mechanics, Faculty of Civil Engineering, Brno University of Technology, Brno 60200, Czech Republic}
\auth[2]{Qiwei  Zhang}
\address[2]{Department of Civil and Environmental Engineering, University of California, Davis, CA, United States}
\auth[2]{John E. Bolander}
\auth[1]{Jan Eliáš\corref{cor1}} 
\cortext[cor1]{Corresponding author, jan.elias@vut.cz}

\begin{abstract}
Two approaches to incorporate heterogeneity in discrete models are compared. In the first, standard approach, the heterogeneity is dictated by geometrical structure of the discrete system. In the second approach, the heterogeneity is imposed by randomizing material parameters of the contacts between the rigid bodies. A~similar randomization strategy is often adopted in continuous homogeneous models.  The study investigates both the elastic and fracture behaviors of these model types, and compares their local and macroscale responses. 

It is found that the stress oscillations present in the standard discrete models built on heterogeneous geometric structures cannot be replicated by randomization of the elastically homogeneous discrete system. The marginal distributions and dependencies between the stress tensor components cannot be adequately matched. Therefore, there is a~fundamental difference between these two views on discrete models. The numerical experiments performed in the paper showed that an~identical response can be achieved at the macroscale by tuning the material parameters. However, the local behavior, fracturing, and internal dependencies are quite different.

These findings provide insight into the potential for controlled random assignment of heterogeneity in homogeneous models. They also demonstrate the need for experimental data capable of verifying the correctness of such an~approach.

\end{abstract}


\end{frontmatter}

\begin{multicols}{2}
\begin{description}
\itemsep -0.5em 
\item[$A$] contact area
\item[$\boldc$]  vector from particle node to integration point
\item[$\boldC$]  covariance matrix
\item[$d$] damage scalar
\item[$\bolde$] strain vector
\item[$\hat\bolde$] vector of eigenstrain
\item[$E$] macroscopic elastic modulus
\item[$E_0$] contact elastic parameter
\item[$\boldf$] external forces
\item[$\feff$] elastic limit in given straining direction
\item[$\ft$] tensile strength at mesoscale
\item[$\Gt$] mesoscale fracture energy in tension
\item[$\hat\boldI$] projection tensor
\item[$l$] contact length
\item[$\boldn$] unit vector of normal reference system 
\item[$\boldt$] traction vector 
\item[$\boldu$] displacement vector 
\item[$V$] rigid-body volume
\item[$\boldx$] coordinates in global ref. system 
\end{description}

\begin{description}
\itemsep -0.5em 
\item[$\alpha$] contact elastic parameter
\item[$\boldvarepsilon$] macroscopic strain vector
\item[$\boldlevicivita$] Levi-Civita tensor
\item[$\varepsilon_V$] volumetric strain
\item[$\boldtheta$] rotation vector
\item[$\boldlambda$] vector of eigenvalues
\item[$\nu$] macroscopic Poisson's ratio 
\item[$\boldxi$] vector of independent random number from standard normal dist.
\item[$\boldsigma$] stress tensor
\item[$\chi$] constitutive history variable
\item[$\boldpsi$] matrix of eigenvectors
\end{description}
\end{multicols}

\section{Introduction}

Discrete models, particularly particle-based lattice models, have been successful in modeling the complex fracture behavior of concrete and other materials with heterogeneous internal structure. There are many types of these models, starting from classical regular and irregular lattice models~\parencite{SchMie92,SchGar97,ZhoAyd-24} and early particle-based models~\parencite{BazTab-90,KikKaw-92,BolSai98,NagSat2005}, up to modern discrete systems with complex constitutive features~\parencite{ChaCif-17,LuoAsa-23,Gra23,DonBri-24}. The most prominent of them is arguably the Lattice Discrete Particle Model (LDPM), which has been successfully used for various structural configurations and loading scenarios~\parencite{CusPel-11,BhaGom-21,ZhuPat-22}. Nowadays, discrete mechanical models are often coupled with other physical or chemical phenomena~\parencite{MiuNak-23,MasKve-23,YinTro-24}.

In contrast to continuum descriptions of material behavior, most particle-based discrete models use vectorial constitutive relations defined at the contacts between neighboring particles, which are assumed to be rigid. The irregular geometry and vectorial constitutive functions give rise to local stress oscillations~\parencite{ZhaEli-24} that can be considered beneficial or unfavorable depending on the goals of the analysis, as described hereafter.

Tessellation of the domain into rigid bodies can be either (i) based on the actual or virtual heterogeneity of the material or (ii) done arbitrarily without consideration of material features. The former approach, called \emph{physical} discretization in Ref.~\parencite{BolEli-21}, may be considered as the standard way of building these models. The stress oscillations are then viewed as an~advantage allowing, for example, to realistically simulate the failure of concrete under  general forms of loading, including compression~\parencite{CusCed07}, which for some applications is an~essential capability.  Such models, however, cannot represent the condition of elastic uniformity and restrict the range of possible values of macroscopic Poisson's ratio~\cite{Eli20}.  Moreover, the stress oscillations arising from the model heterogeneity have never been compared against any experimental data. The notion that they represent 
actual material behavior is yet to be validated.

The latter, \emph{non-physical} tessellation concept is  not related to any material structure. It is understood as an~artificial, purely numerical discretization. Therefore, any stress oscillations due to mesh geometry are inappropriate. Two main approaches, the volumetric-deviatoric decomposition of the constitutive functions~\parencite{CusRez-17} and the auxiliary stress projection method~\parencite{AsaIto-15,AsaKaz-17}, have been developed to eliminate the stress heterogeneity and accurately control the macroscopic Poisson's ratio.  Both approaches exploit information from the neighborhood of a~given nodal site, differing from classical lattice models that involve only two node interactions. The resulting material is elastically homogeneous, which can be beneficial when modeling individual phases or fine-grained matrices of cement-based composite materials~\parencite{KanBol17}, for example. However, the complexity of the model response and its ability to represent heterogeneous media are impaired. The material heterogeneity can be introduced by some form of randomization which can be advantageously controlled independently of the discretization strategy.  This approach has been used in conjunction with finite element analyses of concrete materials and structures, where property variations are introduced on an~elemental basis or at each integration point~\parencite{Breysse90}. With respect to particle-based lattice models, heterogeneity in the form of stiffness variations was introduced for elastic stress analysis~\parencite{ZhaEli-24}. With sufficient grid resolution, the resulting stress fields are independent of the grid size and geometry.  The present article extends this research to different forms of randomization, their properties, and potential implications for both elastic and fracture analyses.
 This latter modeling approach is identical to strategies employed in continuous homogeneous models that use randomization as a~way to incorporate heterogeneous nature of the material.

The two modeling approaches are conceptually very different and, as will be shown in the present paper, provide very different local results. Nevertheless, they exhibit identical, or at least similar, macroscopic behaviors so far. Essential questions arise:
\begin{itemize}
\item What are the differences embedded in these representations of heterogeneity both in elastic and fracture domains?
\item Is it possible to control the imposed randomization in such a~way that the stress oscillations correspond to those from the former approach?   
\item If not, which of the approaches is closer to reality?
\end{itemize}
The present paper elaborates on the first two questions only. The third question is not answered because decisive experimental data is not available. Comparisons with physical test results can follow when techniques for measuring the response quantities of interest, e.g., stress variations at the material scale, become accessible, and the corresponding data are sufficiently available.

Answering these questions is not interesting only for discrete model users. As already mentioned, spatial randomization is also used in homogeneous continuous models to bring some form of material heterogeneity. Learning about correctness of such an~approach in discrete models can be extended to finite element calculations.

\section{Modeling framework}

\begin{figure}[tb!]
\centering
\begin{subfigure}{0.3\textwidth}
\centering
\includegraphics[height=2.8cm]
{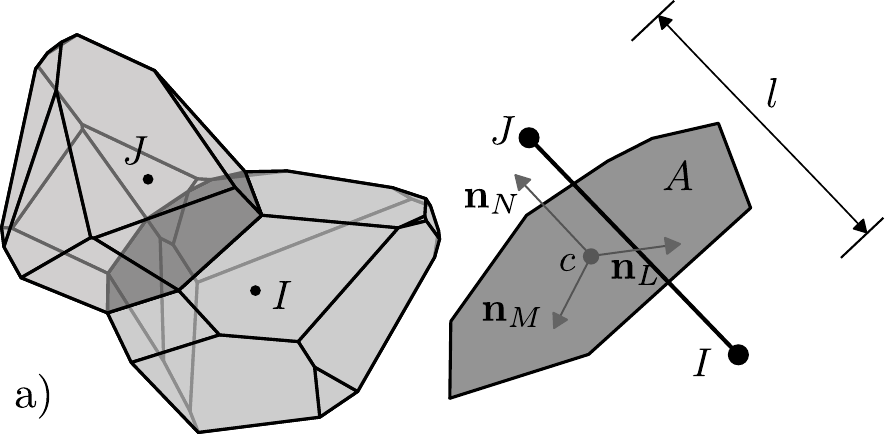}
\end{subfigure}
\centering
\begin{subfigure}{0.68\textwidth}
\centering
\includegraphics[height=4.1cm]{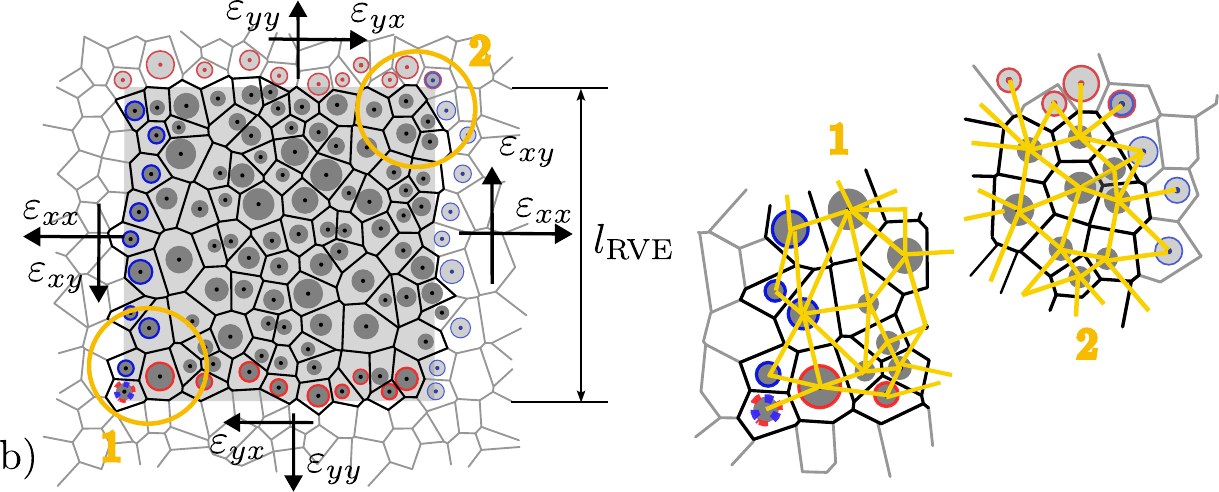}
\end{subfigure}
\caption{a) Two rigid bodies connected by element of length $l$, area $A$, and local reference system $\boldn_{\alpha}$, $\alpha\in\{N,M,L\}$; b) 2D periodic RVE of size $\lRVE$ and boundary nodes of adjacent RVEs; b1) corner with boundary nodes inside RVE and elements (yellow); b2) corner with boundary nodes of adjacent RVEs included in the calculation}
\label{fig:model}
\end{figure}

The three-dimensional discrete models considered herein are built at the meso-scale, where the aggregate inclusions and cementitious binder are viewed as distinct phases of the composite material. Spherical aggregates are generated according to a~specified volume fraction and size distribution, based on a~Fuller sieve curve~\parencite{YanTro-24}. The aggregates are then randomly placed into the domain, starting from the largest, and prohibiting any overlaps. The modified Delaunay triangulation and power (or Laguerre) tessellations~\parencite{OkaBoo-00} are then generated on the aggregate centroids. The Delaunay simplices provide an~estimation of volumetric deformation, $\varepsilon_V$, which is one third of the change in the volume of the simplex~\parencite{CusPel-11}. The power tessellation constitutes rigid bodies that represent aggregates and the surrounding matrix, as shown in Fig.~\ref{fig:model}.

The contacts between rigid bodies are called elements hereinafter, and it will be assumed that they always connect the nodes $I$ and $J$. They have area $A$ and length $l$, the power tessellation ensures the contact plane is perpendicular to the branch vector connecting nodes $I$ and $J$. The local reference system is given by orthonormal vectors $\boldn_N$ (perpendicular to the contact plane), $\boldn_M$, and $\boldn_L$.

The degrees of freedom of the model are the components of displacement, $\boldu$, and rotation, $\boldtheta$, vectors defined at each node $I$. Assuming small displacement and rotations, the strain vector for each element is given by the following kinematic equation
\begin{align}
e_{\alpha} = \frac{1}{l}\left[ \boldu_J - \boldu_I + \boldlevicivita:\left( \boldtheta_J\otimes\boldc_J - \boldtheta_I\otimes\boldc_I \right) \right]\cdot \boldn_{\alpha}
\end{align}
where $\levicivita_{ijk}$ is the Levi-Civita permutation tensor of third order, index $\alpha\in N,M,L$ refers to the directions in the local reference system, and $\boldc_I$ is a~vector pointing from node $I$ to the point $\boldc$ (see Fig.~\ref{fig:model}) where the equation is evaluated. Hereafter, there will be only one such integration point per facet, located at the facet centroid.

The balances of linear and angular momentum for each body $I$ in steady state are expressed as
\begin{align}
\sum_{\alpha} \sum_{e\in I} A_e t^e_{\alpha} \boldn^e_{\alpha} &= \bm{0} &
\sum_{\alpha} \sum_{e\in I} A_e t^e_{\alpha} \boldlevicivita: \boldc^e_I \otimes \boldn^e_{\alpha} &= \bm{0}
\end{align}
The first of these equations sums tractions, $\boldt$, acting at each element $e$ attached to node $I$; the second equation sums moments of those tractions with respect to node $I$. In the presence of body forces, body couples, couple tractions or in a~transient regime, additional terms must be included.

The modeling framework is completed by a~constitutive equation. Several variants of the constitutive model are examined in this paper, as described in the following sections.

\subsection{Elastic constitutive equations}
{\bf Variant \matS}: The standard constitutive model has two material parameters, $E_0$ and $\alpha$. The values used throughout the paper are $E_0=40$\,MPa and $\alpha=0.24$. The traction components reads
\begin{align}
t_N &= E_0 e_N & t_M &= \alpha E_0 e_M & t_L &= \alpha E_0 e_L \label{eq:const_S}
\end{align}
The $\alpha$ parameter is responsible for macroscopic Poisson's ratio, see, e.g., Refs.~\parencite{BatRot88,KuhDadd-00}. This is the most commonly used constitutive model, which results in local oscillations of macroscopic stress in the model~\parencite{ZhaEli-24}. Hereinafter, we will denote this standard constitutive model by letter \matS. The \emph{approximate} relations of $E_0$ and $\alpha$ to the macroscopic elastic parameters $E$ (Young's modulus) and $\nu$ (Poisson's ratio), derived from Voigt's kinematic assumption, are
\begin{align}
\nu &\approx \frac{1-\alpha}{4+\alpha} & E &\approx E_0\frac{2+3\alpha}{4+\alpha}
\end{align}

{\bf Variant \matV}: To have better control over Poisson's ratio and to avoid the stress oscillations, \textcite{CusRez-17} developed a~constitutive model featuring volumetric-deviatoric decomposition and material elastic constants $E_V$ and $E_D$ in constitutive relations
\begin{align}
t_N &= E_D e_N + \left( E_V  - E_D \right) \varepsilon_V & t_M &= \alpha E_D e_M & t_L &= E_D e_L
\label{eq:const_V}
\end{align}
Each contact has several simplices attached to it (as many as the number of vertices of the contact facet), so the volumetric strain  $\varepsilon_V$ is obtained as an~average from  those attached simplices. Because the element formulation in our numerical implementation involves only the degrees of freedom of the two nodes, $I$ and $J$, the volumetric strain must be supplied from outside and the contact stiffness matrix does not have the effect of the volumetric strain included. This means that, even for linear elasticity, this implementation of the constitutive model requires iteration toward meeting the convergence criteria in each time step. This constitutive model will be denoted as \matV in the following text.

The \emph{exact} relations between the $E_D$ and $E_V$ material parameters and macroscopic elastic modulus and Poisson's ratio read~\parencite{CusRez-17}
\begin{subequations}
\begin{align}
E &= \dfrac{3E_V E_D}{E_D+2E_V} & \nu&=\dfrac{E_V- E_D}{E_D+2E_V} \label{eq:EnuVolDev}
\\
E_D&=\frac{E}{1+\nu} & E_V &= \frac{E}{1-2\nu}  \label{eq:EDEVVolDev}
\end{align}
\end{subequations}

{\bf Variant \matH}: The third constitutive model, developed in Refs.~\parencite{AsaIto-15,AsaKaz-17}, follows the standard model from Eq.~\eqref{eq:const_S} with $\alpha=1$. The parameter $E_0$ is then directly equal to the macroscopic elastic modulus (i.e., $E_0 = E$) and the macroscopic Poisson's ratio is 0. Moreover, by setting $\alpha=1$, the discrete model does not exhibit any stress oscillation. In other words, the model is elastically homogeneous, as shown by the proof in Refs.~\parencite{Eli17,Eli20,ZhaEli-24}. The macroscopic stress tensor can be estimated in each rigid body $I$ by the Love-Weber formula~\parencite{Love1927,Web66,DreDeJ72,KryRot96,Bag96,EliCus25}
\begin{equation}
\boldsigma_I =  \frac{1}{V_I} \sum_f  \boldx_f \otimes \boldf_f = \frac{1}{V_I} \sum_{\alpha} \sum_{e\in I} A_e t_{\alpha }\boldc_{I}^e \otimes  \boldn^e_\alpha \label{eq:macrostress}
\end{equation}
where the first summation runs over all external forces $\boldf_f$ acting at position $\boldx_f$, and the second summation acknowledges that these external forces are associated with adjoining elements $e$ for which the contact force becomes $\boldf_f=A_e t_{\alpha} \boldn_{\alpha}$, and $V_I$ is volume of the rigid body. The stress tensor is symmetric unless a~volume couple load or couple traction at the contact facets are applied~\parencite{NicHad-13,YanReg19}.

After evaluating the macroscopic stress for each particle, the average tensorial stress in the contact element $I\!J$ is calculated as $\bar{\boldsigma}_{I\!J}=0.5(\boldsigma_I+\boldsigma_J)$. Its eigen decomposition reads $\bar{\boldsigma}_{I\!J}=\boldpsi_{I\!J}\mathrm{diag}(\boldlambda_{I\!J})\boldpsi_{I\!J}^T$ with eigenvectors collected in columns of the tensor $\boldpsi_{I\!J}$ and principal stresses in vector $\boldlambda_{I\!J}$, the $\text{diag}(\bullet)$ operation transforms the vector into a~diagonal matrix. This decomposition is projected into eigenstrains $\hat \bolde_{I\!J}$ of the element using a~projection tensor $\hat\boldI$~\parencite{AsaIto-15,AsaKaz-17}
\begin{align}
\hat e^{I\!J}_{\alpha} &= \frac{\nu}{E} \boldn^{I\!J}_{\alpha} \cdot \boldpsi_{I\!J} \cdot \mathrm{diag} \left( \hat\boldI \cdot \boldlambda_{I\!J}\right)  \cdot \boldpsi_{I\!J}^T \cdot \boldn^{I\!J}_{N} & \text{where}\quad \hat\boldI &= \left(\begin{array}{ccc}
    0 & 1 & 1 \\
    1 & 0 & 1 \\
    1 & 1 & 0 \end{array}\right) \label{eq:eigenstrain}
\end{align}
The eigenstrains modify tractions, which enter Eq.~\eqref{eq:eigenstrain}. An~iterative loop searching for global balance is applied until the convergence criterion is met. The resulting solution exhibits macroscopic Poisson's ratio and Young's modulus \emph{exactly} equal to $\nu$ and $E$ inserted in Eq.~\eqref{eq:eigenstrain}.

Notice that the \matV and \matH variants produce essentially the same elastically homogeneous model, differing only in the parameters controlling them. In both cases, the resulting medium locally and globally corresponds to a~homogeneous material with identical macroscopic parameters. 

\subsection{Inelastic constitutive equations \label{sec:inelastic}}

The elastic constitutive models in variants \matS and \matH are extended into inelastic behavior
using the damage parameter $d$
\begin{align}  
t_N &=  (1-d) E_0 \left(e_N-\hat e_N\right) & t_M &=  (1-d) E_0 \alpha \left(e_M-\hat e_M\right)& t_L &=  (1-d) E_0 \left(e_L-\hat e_L\right)
\end{align} 
where eigenstrain $\hat e_{\alpha}$ is given by~Eq.~\eqref{eq:eigenstrain} for variant \matH and is zero for variant \matS.

The \emph{non-decreasing} damage parameter is evaluated in space of equivalent strain, $\eeff$, and equivalent traction, $\teff$, as
\begin{align}  
d &= 1-\frac{\teff}{E_0\eeff} 
\end{align}
where
\begin{align}
\eeff & = \sqrt{e_N^2+\alpha e_T^2}  & e_T & = \sqrt{e_M^2+ e_L^2}  &
\teff & = \feff \exp\left( \frac{K}{\feff}\left\langle \chi -\frac{\feff}{E_0}\right\rangle\right)
\end{align} 
The angled brackets return the positive part of the argument. The direction-dependent strength, $\feff$, initial slope of softening/hardening, $K$, and history variable, $\chi$, are
\begin{subequations}
\begin{align}
\feff &= \begin{cases} \dfrac{16\ft}{\sqrt{\sin^2\omega+\alpha\cos^2\omega}} & \omega<\omega_0 \\[5mm] \ft \dfrac{4.52 \sin\omega-\sqrt{20.07 \sin^2\omega+9 \alpha\cos^2\omega }}{0.04\sin^2\omega-\alpha \cos^2\omega} & \omega\geq\omega_0 \end{cases} \label{eq:feq}
\\
K &= \begin{cases} 0.26E_0\left[ 1-\left( \dfrac{\omega+\pi/2}{\omega_0+\pi/2}\right)^2\right] & \omega<\omega_0 \\[5mm] -K_{\mathrm{t}}\left[ 1-\left( \dfrac{\omega-\pi/2}{\omega_0-\pi/2}\right)^{n_t}\right] & \omega\geq\omega_0 
\end{cases}
\\
\chi &= \begin{cases} \eeff & \omega<\omega_0 \\ \eeff\frac{\omega}{\omega_0} + \sqrt{\max e_N^2 + \alpha\max e_T^2}\left(1-\frac{\omega}{\omega_0}\right) & \omega_0\leq\omega<0 \\[1mm] \sqrt{\max e_N^2 + \alpha\max e_T^2} & \omega\geq 0
\end{cases}
\end{align} 
\end{subequations}
Parameter $\omega$ expresses straining direction, $\tan \omega=e_N / \sqrt{\alpha} e_T$, and $\omega_0$ is the direction at which both branches of Eq.~\eqref{eq:feq} are equal. History variables $\max e^2_N$ and $\max e^2_T$ represent maximum square normal and shear strain achieved from the beginning of the simulation. The initial slope of the strain softening, $K$, is defined using $K_{\mathrm{t}}$ and $K_{\mathrm{s}}$, the slopes for pure tension and shear, respectively. These are dependent on contact length $l$ and read
\begin{align}
K_{\mathrm{t}} &= \frac{2E_0\ft^2 l}{2E_0\Gt - \ft^2l} & K_{\mathrm{s}} &= \frac{18\alpha E_0\ft^2 l}{32\alpha E_0\Gt - 9\ft^2l}
\end{align}
Finally, parameter $n_t$ reads
\begin{equation}
n_t = \frac{\ln\left(K_{\mathrm{t}}/(K_{\mathrm{t}}-K_{\mathrm{s}})\right)}{\ln\left( 1-2\omega_0/\pi\right)}
\end{equation}

The presented constitutive model is based on a~formulation developed in Ref.~\parencite{CusCed07}, which was a~predecessor of the Lattice-Discrete Particle Model~\parencite{CusPel-11,AlnPel-19}. It is largely modified by omitting the confinement effect, using damage parameter in shear and compression, and reducing number of material parameters to four: normal elastic constant $E_0$, tangential/normal stiffness ratio $\alpha$, tensile strength $\ft$ and tensile fracture energy $\Gt$. The model has been validated using experimental data from cases dominated by tensile fracture~\parencite{EliVor-15,Eli16,EliVor20}. It has not been optimized and validated for compression.  
 
\subsection{Model geometry and periodic boundary conditions} \label{sec:geometry_materials}

\begin{figure}[tb!]
\centering
\begin{subfigure}{0.32\textwidth}
\centering
\includegraphics[height=3.8cm]
{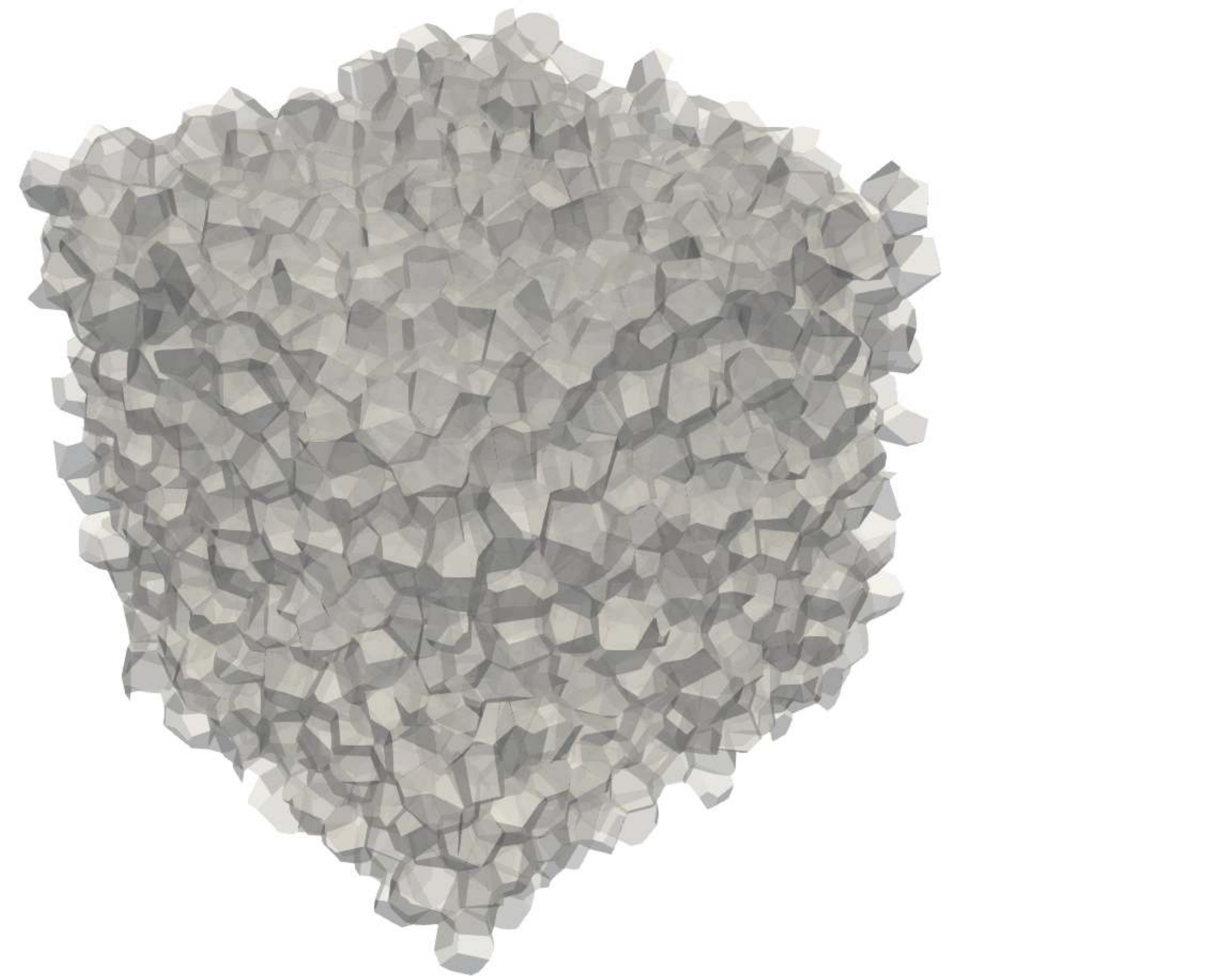}
\end{subfigure}
\begin{subfigure}{0.32
\textwidth}
\centering
\includegraphics[height=3.8cm]{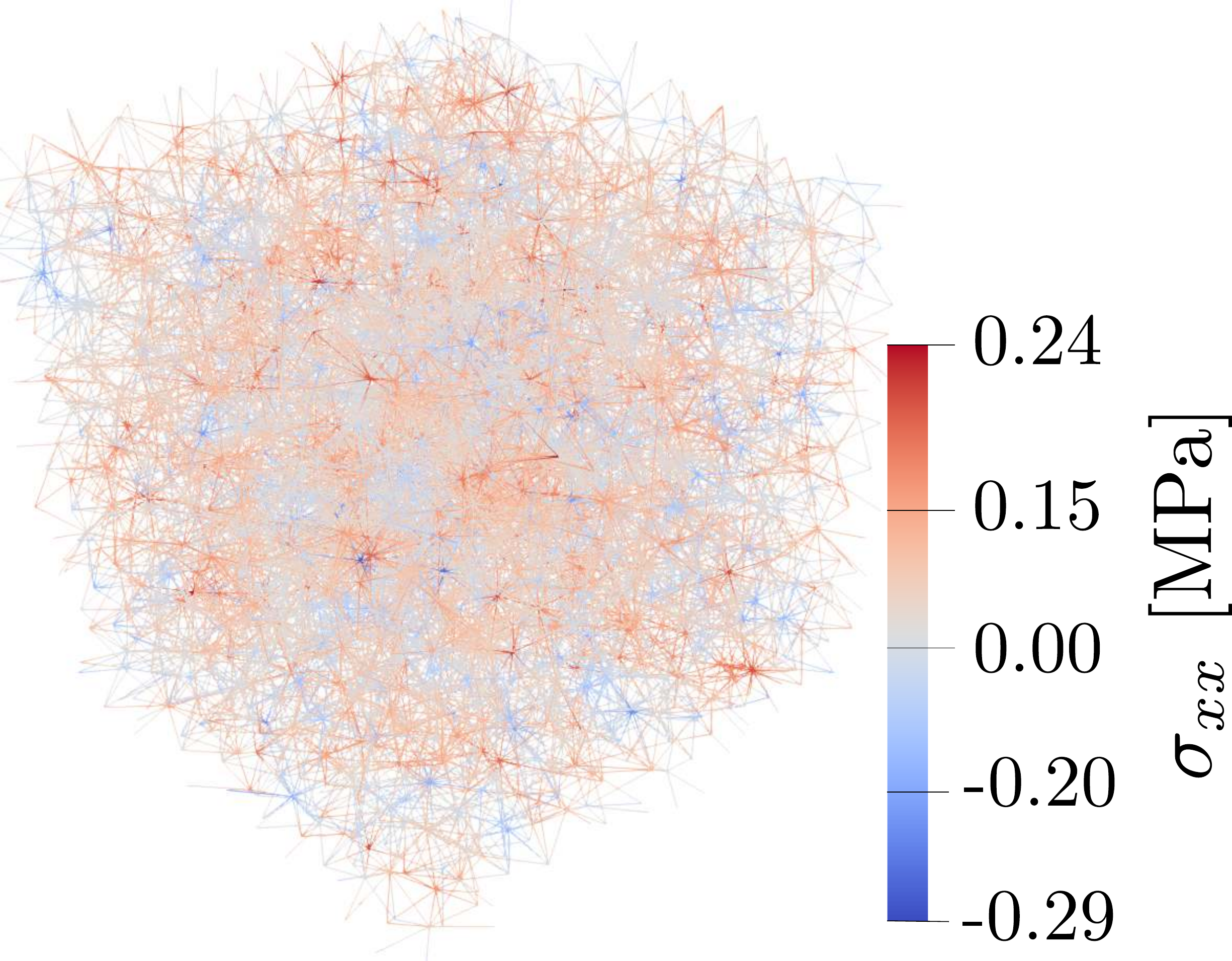}
\end{subfigure}
\begin{subfigure}{0.32
\textwidth}
\centering
\includegraphics[height=3.8cm]{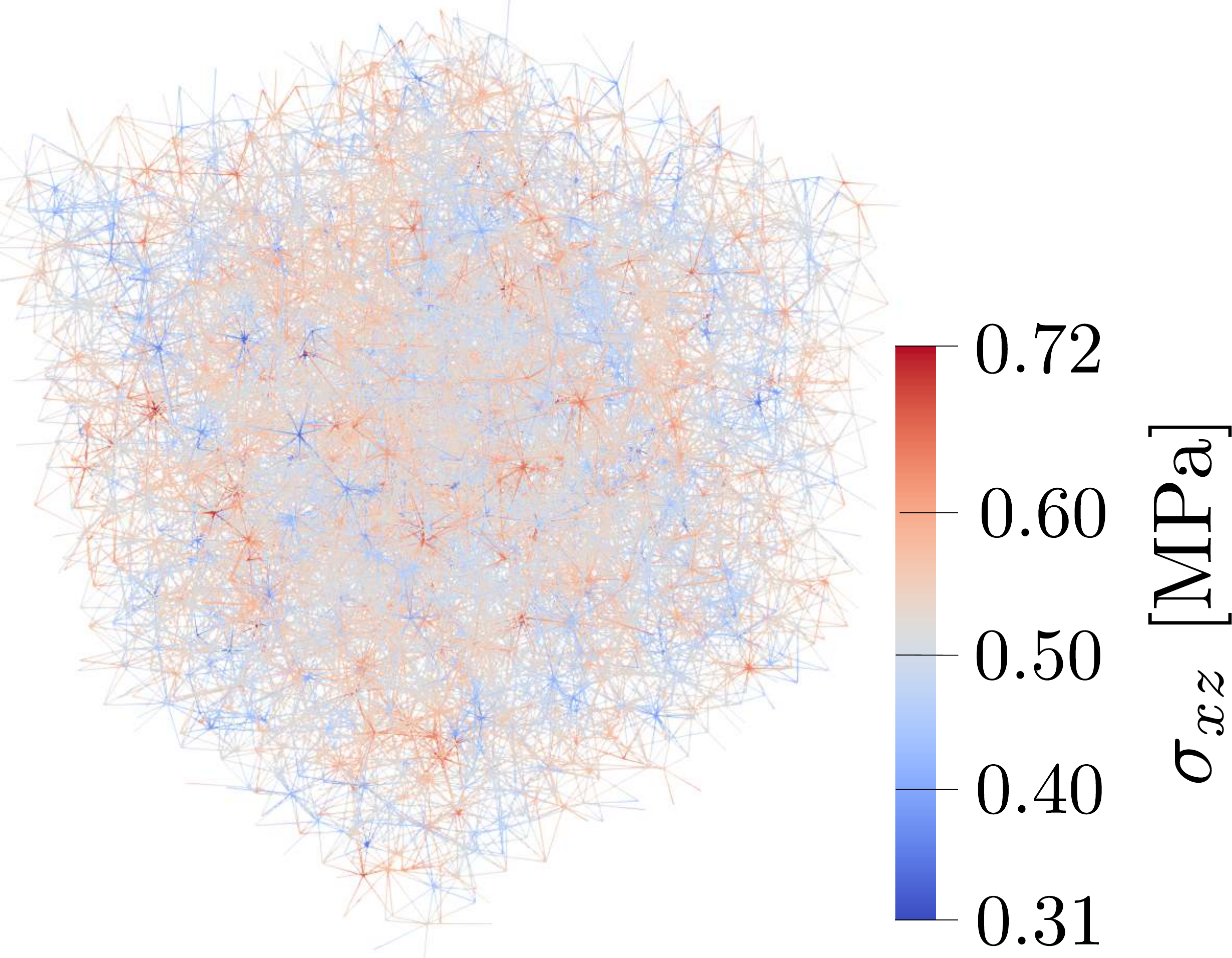}
\end{subfigure}
\caption{a)  periodic set of contact facets in a~3D~RVE of size $\lRVE=100$\,mm; b) discrete elements of the same RVE colored by $\sigma_{xx}$ component of nodal stress under pure shear strain load ($\gamma_{xz} \neq0$); c)  elements of the same RVE colored by $\sigma_{xz}$ under the same load}
\label{fig:100model}
\end{figure}

The model is generated with periodic geometry in a~cube of size $\lRVE$, which is hereinafter called the Representative Volume Element (RVE). Firstly, the Fuller curve is used to generate spheres of diameters within the range 4 to 10\,mm~\parencite{YanTro-24}. These spheres are placed randomly into the domain without overlapping, starting from the largest diameter spheres. The Euclidean metric used to check whether the overlapping is periodic,  i.e., distance between two points $I$ and $J$ along the axis $i$ is $\Delta x_i = \min(|x^J_i-x^J_i|,0.5\lRVE-|x^J_i-x^J_i|)$~\parencite{EliVor16,VorEli20}. Then, periodic power tessellation is performed.  The periodicity is enforced by periodic repetition of the particles close to the domain boundary, see, e.g., Ref.~\parencite{EliVor-20} where the same strategy is used for Voronoi tesselation. The main advantage of the fully periodic geometry is complete elimination of all boundary effects described in Ref.~\parencite{Eli17}. An~example of a~2D~periodic RVE is presented in Fig.~\ref{fig:model}b.

The boundary conditions applied to the model are the following. There is a~prescribed macroscopic strain tensor, $\boldvarepsilon$, that drives the difference of displacements of the primary ($I$) and dependent ($J$) nodes at the opposite surfaces of the periodic structure
\begin{align}
\boldu_J &= \boldu_I + \left(\boldx_J-\boldx_I \right)\cdot\boldvarepsilon &
\boldtheta_J &= \boldtheta_I \label{eq:periodicload}
\end{align}

In Fig.~\ref{fig:model}b, the pairs of primary and dependent nodes are highlighted with red and blue colors for the top-bottom and left-right boundaries, respectively. Additionally, the lower left and upper right nodes are also dependent. Owing to the periodic geometry, elements are not used to connect nodes on two of the four boundaries (here the upper and right boundaries) to prevent element duplication, as illustrated in Fig.~\ref{fig:model}b2.
To eliminate rigid-body translations, one node is randomly selected and its displacements are restricted. Rigid-body rotations are automatically prevented by the period boundary conditions. The applied loading differs from the one derived by asymptotic homogenization in Refs.~\parencite{RezCus16,RezZho-17,EliCus22}, which features projection of the macroscopic strain tensor into eigenstrains in discrete contact elements. The loading through eigenstrains is replaced for the sake of simplicity because the eigenstrains are already used to introduce the Poisson's effect in Eq.~\eqref{eq:eigenstrain}. If needed, both eigenstrain sources can be used simultaneously, but the direct loading through Eq.~\eqref{eq:periodicload} provides exactly identical results and its description in connection with Eq.~\eqref{eq:eigenstrain} seems less complicated.

3D periodic models of size $\lRVE = 100$\,mm are used for the elastic analysis and size $\lRVE=50$\,mm for the inelastic analysis. A~set of 100 models differing by internal geometry is generated for each RVE size. On average, each model of size $\lRVE=100$\,mm contains about 19900 elements and 3400 nodes. In inelastic analyses, stress field oscillations are not inspected; only the global response character is examined. The smaller size models were used with the advantage of significantly faster computation times.

All simulations are performed in the open source software OAS (Open Academic Solver) available at \url{https://gitlab.com/kelidas/OAS}. This software is developed at Brno University of Technology as a~multipurpose platform for the solution of various physical problems. The simulations are computed as steady state using a~modified Newton-Raphson iteration scheme. 

\section{Linear elastic material behavior}
The first investigation is directed towards elastic behavior. The following material variants,  summarized in Table~\ref{tab:variants}, and parameters are used.
\begin{itemize}
\item {\bf Variant \matS}, the standard material with material parameters $E_{0} = 40$\,GPa and $\alpha = 0.24$, values commonly used for ordinary concrete. For further analysis, the effective macroscopic elastic modulus $\effE$, and the effective macroscopic Poisson's ratio, $\effnu$, are found for each RVE. An~effective macroscopic stiffness tensor is constructed by loading each RVE by strain in six linearly independent directions (three normal and three shear loads). The effective values of $E$ and $\nu$ are found by numerical fitting of the theoretical isotropic stiffness tensor to this computed tensor.

\begin{table}[tb!]
\centering

\begin{tabular}{c c c c c c} \toprule
variant & subvariant  & model & homogenization& discretization  & randomized \\ 
 & & form & method & scheme & parameters \\  \midrule \midrule
\sf{S} & - & standard  & none & physical & - \\ \midrule
\sf{V} & - & homogeneous  & volumetric-deviatoric split& non-physical & - \\
  & \sf{RD} & & & non-physical & $E_D$\\ \midrule
\sf{H} & - & homogeneous & tensorial stress projection &non-physical & - \\
  & \sf{R} & &  &non-physical & $E,\,\nu$ \\
  & \sf{RD} & & &non-physical & $E_D$ \\
  & \sf{RF} & &  &non-physical & rand. field $E,\,\nu$
   \\\bottomrule
\end{tabular}

\caption{Model variants used for elastic stress analysis, the \matHRF\ material model uses stationary random field to assign random material properties, other randomized models sample each location independently.}
\label{tab:variants}

\end{table}

The choice of the elastic modulus is not important, it only linearly scales the results. However, the value of $\alpha$ affects the stress oscillations. For $\alpha=1$, the stress oscillations completely disappear; see, e.g., Ref.~\parencite{Eli20,ZhaEli-24}.

\item {\bf Variant \matV} with the volumetric-deviatoric split uses the effective parameters from the variant \matS in Eq.~\eqref{eq:EDEVVolDev} for each internal geometry of RVE: $E_D = \effE/\left(1+\effnu\right)$ and $E_V=\effE/\left(1-2\effnu\right)$. The randomized version of this material variant applies randomness only to the deviatoric part.
\begin{itemize}
\item {\bf Subvariant \matVRD} has parameter $E_D$ randomly drawn from a~uniform distribution with user-defined width. The width is specified by the parameter $\eta$ that determines the lower bound as $(1-\eta)\mu_D$ and the upper bound as $(1+\eta)\mu_D$ where $\mu_D$ is the mean value. The mean value of the uniform distribution is set so that each RVE stiffness with \matVRD material matches the stiffness of the same RVE geometry with \matS material, i.e., each RVE geometry uses a~distribution with slightly different mean value. The reason why the randomization changes the mean value, apart from the insufficient statistical sample, is simply the interaction between elastic contacts. In an~extreme case of all contacts connected in series, the elastic modulus would read $E=\left(\sum_i 1/ E_i\right)^{-1}$. In another extreme case of parallel connectivity, the elastic modulus would read $E=\sum_i E_i$. The complex structure of the RVE gives values between these extremes. Both the mean of $E_D$ and the deterministic value of $E_V$ are modified so that the \matVRD RVE stiffness tensor matches the \matS RVE stiffness tensor.
\end{itemize}  

\item {\bf Variant \matH}, a~homogeneous material based on the projection of tensorial stresses, with parameters $E=\effE$ and $\nu=\effnu$. These parameters are, as in the case of the variant \matV, different for each RVE geometry, so the stiffness of each RVE with material \matH matches the stiffness calculated for material \matS. These elastic parameters for each element are also randomized by one of the following approaches.
\begin{itemize}
\item {\bf Subvariant \matHR} assumes both $E$ and $\nu$ uniformly distributed and independent, and without any spatial correlation. In other words, $E$ and $\nu$ for each element are drawn independently from uniform distributions with mean values $\mu_E$ and $\mu_{\nu}$ and width parameters $\eta_E$ and $\eta_{\nu}$. The meaning of the width parameters is identical to the one in \matVRD variant and they are specified by the model user. Since the mean values for each randomized RVE are computed in the same way as for \matVRD variant, the stiffness tensor must match the stiffness tensor of the RVE with standard (\matS) material.

\item {\bf Subvariant \matHRD} randomizes only the deviatoric component of the material stiffness, $E_D$, while keeping the volumetric part constant. For each element the parameter $E_{D}$ is drawn from a~uniform distribution and values of elastic modulus and Poisson's ratio are then calculated from Eq.~\eqref{eq:EnuVolDev}. The width parameter $\eta_D$ is a~free model parameter while its mean value and constant $E_V$ are updated as in all the previous random models so stiffness of each randomized RVE corresponds to stiffness of the RVE with the standard (\matS) material. 

\item {\bf Subvariant \matHRF} employs independent $E$ and $\nu$ generated as realizations of random fields with uniform marginal distributions. The uniform distribution has user-defined widths (through $\eta$ parameters) and mean values correspond to the mean effective values from all 100 RVEs with the standard (\matS) material. The Gaussian (square exponential) autocorrelation function is used and the correlation length $\lcorr$ is another user-defined material parameter. 

Random fields are generated via the Karhunen--Lo\`{e}ve expansion~\parencite{Kar49,Spa89,Ste09,Gha03}. It is based on the spectral decomposition of covariance matrix $\boldC$, which transforms the correlated Gaussian variables into independent standard Gaussian variables, which are simple to generate. The covariance matrix $\boldC$ is computed with the help of the periodic metric~\parencite{EliVor16,VorEli20} so the resulting field is periodic as well. The periodic Gaussian random field is obtained using vector $\boldxi$ of $K$ standard Gaussian random variables
\begin{align}
H(\boldx) = \displaystyle \sum_{k=1}^K\sqrt{\lambda_k} \xi_k
\boldpsi_k(\boldx) \label{eq:KLexpansion}
\end{align}
where $\lambda$ and $\boldpsi$ are the eigenvalues and the eigenvectors of the covariance matrix $\boldC$, and $K$ is the number of eigenmodes considered. Only $K$ eigenmodes corresponding to $K$ largest eigenvalues are used, so that $\sum_{k=1}^K \lambda_k$  is at least 99\,\% of the trace of the covariance matrix $\boldC$ \parencite{Vor08}.  The independent standard Gaussian variables $\boldxi$ must be sampled carefully. Ideally, one should use Latin Hypercube Sampling and some robust optimization strategy allowing optimal integration properties of the resulting set \parencite{VorNov09,EliVor16,VorEli20}.

The Gaussian field is generated on a~grid and then projected onto the discrete mechanical structure by the expansion-optimal linear estimation method (EOLE) developed by \textcite{LiKiu93}), see papers~\parencite{EliVor-15,EliVor20} which apply the identical strategy. 

After the Gaussian random field values at facet centers are obtained, they are transformed to non-Gaussian space using isoprobabilistic transformation. This transformation distorts the correlations, therefore when generating the underlying Gaussian field in Eq.~\eqref{eq:KLexpansion}, the correlation coefficients in matrix $\boldC$ must be modified in order to fulfill the desired pairwise correlations of the non-Gaussian field. This is provided by the Nataf model, which can be
evaluated by an~approximation from Ref.~\parencite{LiLu-08}.
\end{itemize}

\end{itemize}

One can randomize the variant \matV in a~different way (using random fields, randomizing also $E_V$ parameter, or randomize some combination of $E_V$ and $E_D$, for example as done in \matHR by independent randomization of $E$ and $\nu$). As discussed earlier, variants \matV and \matH are essentially identical, differing only in the way macroscopic material properties are assigned. The other randomization variants of \matV material would therefore only revisit the \matH material randomization. However, in the \matH variant the application of the spatial randomness is blended with computing the average stress tensor in the rigid bodies. In the \matV variant, it is applied directly to the element. That is the primary reason why variant \matVRD is introduced. It can be directly compared to \matHRD variant to observe how different forms of introducing the same form of randomness affect the results.

In all the random models only uniform distribution is used. Other distributions can be easily inserted instead. Also, one can study the effect of the relationship between random values of the two elastic parameters, so-called cross correlation. These options are not investigated in the present study but shall be addressed in the future.

\begin{table}[tb!]
\centering
\begin{tabular}{l l l c c} \toprule
load & $\{\varepsilon_{xx}, \varepsilon_{yy}, \varepsilon_{zz}, \varepsilon_{yz}, \varepsilon_{xz}, \varepsilon_{xy}\} \  (\times 10^{-5})$  & $\boldvarepsilon_\mathrm{dev} \  (\times 10^{-5})$ & $\varepsilon_{V} \  (\times 10^{-5})$ &  $\varepsilon_{\mathrm{eq}} \  (\times 10^{-5})$     \\ \midrule
1 & $\{1.5, 1.5 , 1.5, 0, 0, 0  \}$ & $\{0, 0, 0, 0, 0,  0  \}$  &  1.5 & $0 $ \\  
2 & $\{0,0 ,0, 1.50, 1.50,   1.50  \}$ & $\{0, 0, 0,  1.50,  1.50,  1.50   \}$  &  0.0 & $4.50 $ \\  
3 & $\{0, 0, 0, 0, 0,  2.60  \}$ & $\{0, 0, 0, 0, 0,  2.60   \}$  &  0.0 & $4.49 $ \\  
4 & $\{1.5, 1.5, 1.5, 0, 0,  2.60  \}$ & $\{0, 0, 0, 0, 0,  2.60   \}$  &  1.5 & $ 4.49 $
 \\\bottomrule
\end{tabular}
\caption{Applied strains for elastic analysis}
\label{tab:volum_load_vector}
\end{table}

The RVEs of size $\lRVE=100$\,mm with different variants of elastic constitutive models are subjected to macroscopic strain loading. Tensorial stresses are evaluated by Eq.~\eqref{eq:macrostress} for the node of each rigid body. Due to perfect periodicity of the whole model, the ergodicity of the stress field is ensured, and statistical sampling in space and over RVE realizations is equivalent. A~statistical assessment is presented for different loading scenarios: (i) volumetric loading, (ii) deviatoric loading, and (iii) general loading composed of both volumetric and deviatoric parts. The strain loading tensors are listed in Tab.~\ref{tab:volum_load_vector}. For each case, 100 RVEs with different internal structures and different random values of material parameters (if applied) are included in the analysis.

\subsection{Volumetric strain load}
The imposed macroscopic strain tensor, denoted \emph{Load 1}, reads $\varepsilon_{ij} = \varepsilon_V \delta_{ij}$, where $\delta_{ij}$ is Kronecker delta and $\varepsilon_V$ is a~scalar representing volumetric strain. $\varepsilon_V$ is chosen to be $1.5 \times 10^{-5}$ without loss of generality, as any other value would produce identical results only multiplied by the appropriate constant. 

\begin{table}[tb!]
\centering
\begin{tabular}{llCCCCCC} \toprule
\multicolumn{2}{l}{material variant} & \sigma_{xx} / \sigma_V & \sigma_{yy} / \sigma_V & \sigma_{zz}/ \sigma_V & \sigma_{yz}/ \sigma_V & \sigma_{xz}/ \sigma_V & \sigma_{xy}/ \sigma_V \\    
\midrule
\multirow{2}{1cm}{\matS} &  average & 1.000 & 1.000 & 1.000 & \sim 10^{-22} & \sim 10^{-22} & \sim 10^{-22}
\\
& stand. dev. & \sim 10^{-20} & \sim 10^{-20} & \sim 10^{-20} & \sim 10^{-20} &  \sim 10^{-20} & \sim 10^{-20} 
\\ \midrule
\multirow{2}{1cm}{\matH}&   average & 1.000 & 1.000 & 1.000 & \sim 10^{-16} & \sim 10^{-16} & \sim 10^{-16} 
\\
& stand. dev. & \sim 10^{-14} & \sim 10^{-14} & \sim 10^{-14} & \sim 10^{-14} &  \sim 10^{-14} & \sim 10^{-14} 
\\ \midrule
\multirow{2}{1cm}{\matHR} & average & 1.000 & 1.000 & 1.002 & \sim 10^{-10} & \sim 10^{-10} & \sim 10^{-10} 
\\
& stand. dev. & 0.105 & 0.105 & 0.103 & 0.053 & 0.053 & 0.053   
\\ \midrule
\multirow{2}{1cm}{\matHRD} & average & 1.000 & 1.000 & 1.000 & \sim 10^{-14} & \sim 10^{-14} & \sim 10^{-14} 
\\  
& stand. dev. & \sim 10^{-13} & \sim 10^{-13} & \sim 10^{-13} & \sim 10^{-14} &  \sim 10^{-14} & \sim 10^{-14}
\\ \midrule
\multirow{2}{1cm}{\matVRD} & average & 1.000 & 1.000 & 1.002 & \sim 10^{-10} & \sim 10^{-10} & \sim 10^{-10} 
\\
& stand. dev. &  \sim 10^{-13} & \sim 10^{-13} & \sim 10^{-13} & \sim 10^{-13} &  \sim 10^{-13} & \sim 10^{-13} 
\\ \bottomrule
\end{tabular}
\caption{Averages and standard deviations of stress field components produced by volumetric \emph{Load 1}, $\varepsilon_V = 1.5 \times 10^{-5}$ acting on 100 RVEs. The average and standard deviation values have been normalized by the theoretical value of volumetric stress $\sigma_{V} = E_V 
 \varepsilon_V = 0.600 \ \text{MPa}$. The width parameters are $\eta_{E} = \eta_{\nu} = 0.95$ for \matHR material and $\eta_{D}=0.8$ for \matHRD and \matVRD materials.}
\label{tab:pure_volumetric}
\end{table}

As documented in Tab.~\ref{tab:pure_volumetric}, the tensorial stress field produced by the standard material (\matS) is constant for all rigid bodies within machine precision. The resulting stress is hydrostatic, the normal components are identical and there are no shear stresses. The standard deviations of all the components are practically zero. The very same is seen for \matH material variant, except this time the variances are higher as the iterative procedure stopped at some threshold and therefore introduced random numerical error into the solution. Lowering the convergence limit would decrease the variances but also demand more computational resources. The results for \matV material variant are not reported in the table but they are identical to the \matH variant. These models are, in the deterministic setup, indistinguishable. 

Further rows in Tab.~\ref{tab:pure_volumetric} show results for the randomized material variants. The selected distribution width parameters are the following:  $\eta_{E} = \eta_{\nu} = 0.95$ for \matHR material and $\eta_{D}=0.8$ for \matHRD and \matVRD materials.  It is seen that randomization of $E_D$ does not change the means and variances, because deviatoric stiffness is not activated under the volumetric loading. Therefore, both \matHRD and \matVRD variants give results identical to deterministic models. This is true not only statistically but also locally, the results for elements within each RVE are identical to the deterministic models. The only difference is seen for the \matHR variant, where both elastic modulus and Poisson's ratio are randomized. The mean values still converge to the mean values from deterministic variants with number of samples, however, the variances are much larger this time and cannot be decreased by improving convergence criteria. The stress solutions over the RVE are not constant but oscillate in space.

\subsection{Deviatoric strain load}

\begin{figure}[tb!]
\centering
\includegraphics[width=2.15in]{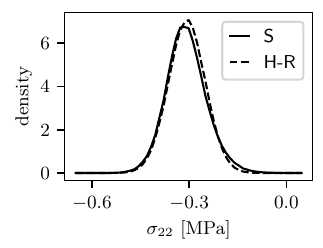}\hfill
\includegraphics[width=2.15in]{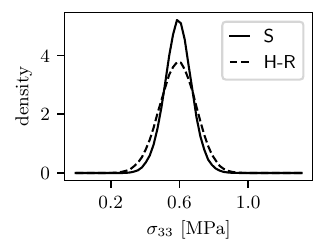}\hfill
\includegraphics[width=2.15in]{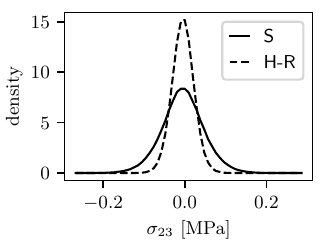}
\caption{Examples of histogram differences $\Delta_{\mathrm{hist}}$ between \matS and \matHR material variants computed on RVEs subjected to \emph{Load 2}: from the left $\Delta_{\mathrm{hist}} = 1.723$; -2.970; 14.234.}
\label{fig:Load1_histograms}
\end{figure}

The RVEs are now subjected to purely deviatoric strain loads. The tensorial shear strains $\varepsilon_{ij} = \gamma_{ij} / 2$ where $  i \neq j$ are prescribed.  In \emph{Load 2} all the shear strains are equal to $1.5\times10^{-5}$ and all normal strains are zero. In \emph{Load 3} all strain components are zero except $\varepsilon_{xy}=1.5\sqrt{3}\times10^{-5}$. The deviatoric strain measure $\varepsilon_{\text{eq}}$ computed equivalently to Von Mises stress is equal to $4.5\times10^{-5}$ for both deviatoric loading scenarios. The stress tensors are studied in the original reference system $xyz$ and their components denoted $\sigma_{xx}$, $\sigma_{yy}$, $\sigma_{zz}$, $\sigma_{yz}$, $\sigma_{xz}$, and $\sigma_{xy}$, but also in rotated reference system corresponding to principal directions of the macroscopic strain tensor where the components are denoted $\sigma_{11}$, $\sigma_{22}$, $\sigma_{33}$, $\sigma_{23}$, $\sigma_{13}$, and $\sigma_{12}$.

The \matH and \matV material variants produce a~homogeneous stress field corresponding to the linearly elastic continuum by design in both loading scenarios. The standard material (\matS) and all the randomized materials exhibit stress oscillations. Histograms of selected principal stress components under \emph{Load 2} are shown in Fig.~\ref{fig:Load1_histograms}. The \matS material model is compared to \matHR models with $\eta_{E} = \eta_{\nu} = 0.7$. This $\eta$ choice produces $\sigma_{22}$ stress oscillations in the \matHR material similar to the \matS material, but $\sigma_{33}$ and $\sigma_{23}$ histograms are wider and narrower, respectively, than the corresponding histograms generated by \matS material. 

For the comparison of the stress fields of the different materials, the $\Delta_{\mathrm{hist}}$ value is introduced. It measures the distance of two histograms of nodal stress components as Euclidean distance $\Delta_{\mathrm{hist}} = \pm \sqrt{\sum_{i = 1}^{n}\left[h_{1}(i) - h_2(i)\right]^2}$ where $h_{1}(i)$ and $h_{2}(i)$ is the frequency of the $i$th  bin in the first and the second histogram, respectively, and $n$ is the number of bins, which is identical ($n=30$) in all histograms. The sign is added to indicate which distribution is wider and which is narrower. A~negative sign means the peak of the first material (usually the \matS variant) is larger and vice versa. The $\Delta_{\mathrm{hist}}$ values of the histogram pairs from Fig.~\ref{fig:Load1_histograms} are stated in the figure caption.

\begin{figure}[tb!]
\centering
\includegraphics[width=6.5in]{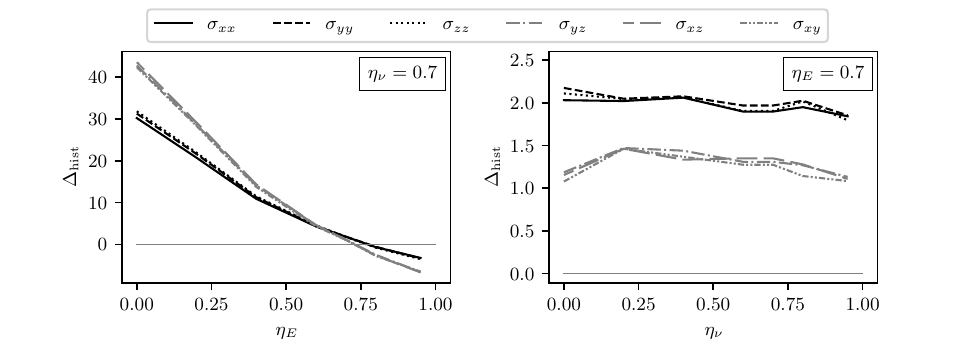}
\includegraphics[width=6.5in]{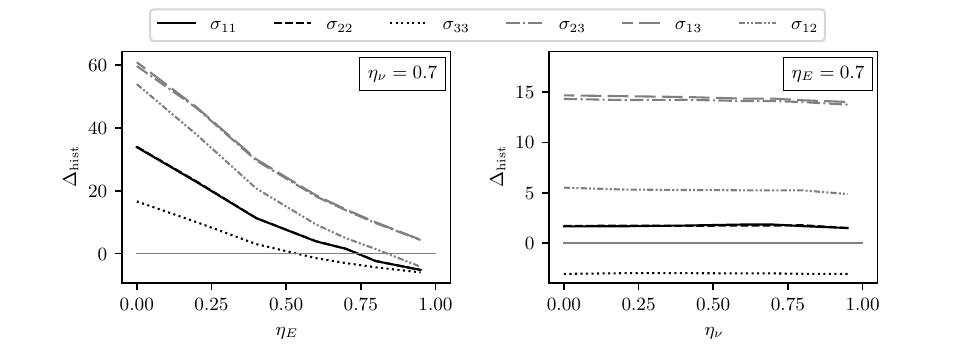}
\caption{Histogram distance $\Delta_{\mathrm{hist}}$ between \matS and \matHR material variants for different $\eta_E$ and $\eta_{\nu}$ parameters of randomization for \emph{Load 2}. Top row: stress components in $xyz$ reference system, bottom row: stress components in the reference system of principal strains, left column: fixed value $\eta_{\nu} = 0.7$, right column: fixed value $\eta_{E} = 0.7$.}
\label{fig:nu_random_metric}
\end{figure}

Firstly the RVEs of \matS and \matHR materials were subjected to \emph{Load 2}. The influence of the $\eta_{E}$ and $\eta_{\nu}$ governing randomization of the elastic parameters was inspected by measuring distance of stress histograms for \matS and \matHR material. Figure~\ref{fig:nu_random_metric} on the right-hand side shows that changing $\eta_{\nu}$ at constant $\eta_{E} = 0.7$ does not change the histogram distance, the histograms of all the components are insensitive to $\eta_{\nu}$. There is, however, strong sensitivity of them to $\eta_{E}$. This is seen in Fig.~\ref{fig:nu_random_metric} on the left-hand side for constant $\eta_{\nu} = 0.7$. Lower values of $\eta_{E}$ produce narrower histograms of stress components. Looking at the stress tensors in the $xyz$ reference system (the top row in Fig.~\ref{fig:nu_random_metric}), one can conclude that for $\eta_{\nu} = \eta_{E} = 0.7$, the stress oscillations in \matHR material closely resemble those in \matS material. Surprisingly, when transformed to the macroscopic principal directions, such an~optimal value cannot be found as documented in the bottom row of Fig.~\ref{fig:nu_random_metric}. Also, the same random settings applied with $\emph{Load 3}$ do not lead to sufficient match, neither for $xyz$ nor $123$ reference system. All the histogram distances for $\eta_{\nu} = \eta_{E} = 0.7$ are listed in Tab.~\ref{tab:load12_metric}.

The dependency of the \matHR histograms on the chosen reference system is attributed to existing relations between stress tensor components. 
Table~\ref{tab:corr_mat} shows Spearman (linear) correlation matrices of standard (\matS) and \matHR material stress field components in reference systems $xyz$. The correlations for the standard material are negligible, while for \matHR material they are significant. When some reference system transformation is applied, the almost independent standard material components transform differently from the \matHR material, which exhibits significant dependencies. The correlations of stress field components were also inspected for an~analogous continuum model. A~periodic RVE with a~similar number of nodes was constructed from three-dimensional trilinear isoparametric finite elements. The elastic parameters ($E$ and $\nu$) of the finite elements were randomized in the same fashion as in the \matHR material and the resulting correlation matrices reveal similar values as in the \matHR material. The discrete \matH material behaves the same way as the continuous elastic model.

\begin{figure}[tb!]
\centering
\includegraphics[width=6.5in]{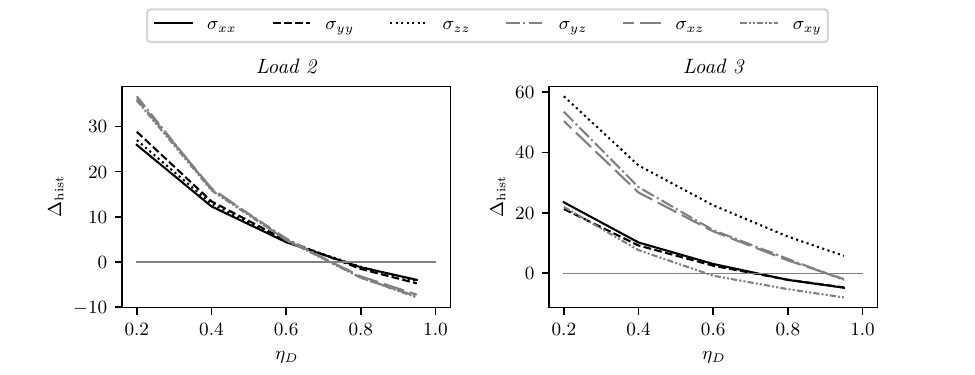}
\caption{Histogram distance $\Delta_{\mathrm{hist}}$ between \matS and \matHRD material variants for different $\eta_D$ parameters of randomization for \emph{Load 2} and \emph{Load 3}}
\label{fig:HRD_random_metric}
\end{figure}

\emph{Load 3} represents a~different deviatoric load type. Attempts to minimize histogram distance $\Delta_{\mathrm{hist}}$ between \matHR and \matS materials by changing $\eta_E$ and $\eta_{\nu}$ ended with unsatisfactory results for both $xyz$ and $123$ reference systems. The lowest values of $\Delta_{\mathrm{hist}}$ found can be seen in Tab.~\ref{tab:load12_metric} as well. Randomizing the $E$ and $\nu$ parameters through $E_D$ in \matHRD material variant leads to similarly poor results as in the case of \matHR material. On the left-hand side of Figure \ref{fig:HRD_random_metric}, it is shown that while it is once again possible to minimize the $\Delta_{\mathrm{hist}}$ for all components in the reference \emph{xyz} system for the \emph{Load 2}, the right-hand side shows that under \emph{Load 3} no $\eta_D$ value leads to oscillations matching those in the \matS material.
Finally, the \matVRD material variant produced another unsatisfactory result: there is no $\eta_D$ value for which the histograms of all the stress components match those from the \matS material. 

\begin{table}[tb!]
\centering
\small
\begin{tabular}{l l C C C C C C} \toprule
ref. system & load & \sigma_{xx} & \sigma_{yy} & \sigma_{zz} & \sigma_{yz} & \sigma_{xz} & \sigma_{xy} \\ \midrule
\multirow{2}{2cm}{coaxial, $xyz$} & \emph{Load 2} & 2.062 & 2.078 & 2.065 & 1.440 & 1.336 & 1.370 \\   & \emph{Load 3} & 1.852 & 1.722 & -3.051 & 12.703 & 12.922 & 4.891  \\
\multirow{2}{2cm}{principal, 123} & \emph{Load 2} & 1.751 & 1.723 & -2.970 & 14.234 & 14.514 & 5.283 \\ & \emph{Load 3} & -3.312 & -3.550 & 18.051 & 9.822 & 9.095 & 18.198  \\ \bottomrule
\end{tabular}
\caption{Histogram distance $\Delta_{\mathrm{hist}}$ between \matS and \matHR ($\eta_{E} = \eta_{\nu} = 0.7$ ) material variants in coaxial and principal directions for \emph{Load 2} and \emph{Load 3}.}
\label{tab:load12_metric}
\end{table}

\begin{table}[tb!]
    \centering
    \begin{tabular}{c}
    standard material (\matS) \\
    $\begin{bNiceArray}{cccccc}[first-row,first-col]
     & \sigma_{xx} & \sigma_{yy} & \sigma_{zz} & \sigma_{yz} & \sigma_{xz} & \sigma_{xy} \\
    \sigma_{xx}& 1.0 & -0.03 & -0.04 & -0.08 & 0.11 & 0.11  \\
    \sigma_{yy}& -0.03 & 1.0 & -0.03 & 0.11 & -0.08 & 0.11  \\
    \sigma_{zz}& -0.04 & -0.03 & 1.0 & 0.11 & 0.11 & -0.08  \\
    \sigma_{yz}& -0.08 & 0.11 & 0.11 & 1.0 & 0.17 & 0.17  \\
    \sigma_{xz}& 0.11 & -0.08 & 0.11 & 0.17 & 1.0 & 0.17  \\
    \sigma_{xy}& 0.11 & 0.11 & -0.08 & 0.17 & 0.17 & 1.0  \\
    \end{bNiceArray}$
    \end{tabular}
    \begin{tabular}{c}
     randomized homogeneous mat. (\matHR) \\
    $\begin{bNiceArray}{cccccc}[first-row,first-col]
     & \sigma_{xx} & \sigma_{yy} & \sigma_{zz} & \sigma_{yz} & \sigma_{xz} & \sigma_{xy} \\
\sigma_{xx}& 1.0 & 0.51 & 0.51 & 0.33 & 0.26 & 0.26  \\
\sigma_{yy}& 0.51 & 1.0 & 0.51 & 0.26 & 0.33 & 0.26  \\
\sigma_{zz}& 0.51 & 0.51 & 1.0 & 0.26 & 0.26 & 0.33  \\
\sigma_{yz}& 0.33 & 0.26 & 0.26 & 1.0 & 0.55 & 0.55  \\
\sigma_{xz}& 0.26 & 0.33 & 0.26 & 0.55 & 1.0 & 0.55  \\
\sigma_{xy}& 0.26 & 0.26 & 0.33 & 0.55 & 0.55 & 1.0  \\
\end{bNiceArray}$
    \end{tabular}
    \caption{Correlations between components of the stress tensor in rigid bodies expressed in $xyz$ reference system under \emph{Load 2}; left: standard material (\matS); right: randomized homogeneous material (\matHR) with $\eta_{E} = \eta_{\nu} = 0.7$.}    
    \label{tab:corr_mat}
      
\end{table}

\begin{figure}[tp!]
\centering
\includegraphics[width=6in]{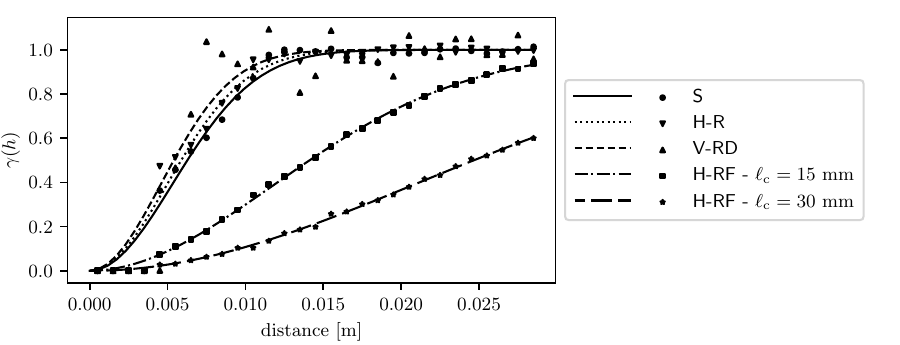}
\caption{Variogram of $\sigma_{xx}$ under \emph{Load 2} for different autocorrelation lengths $\lcorr$.}
\label{fig:vario_0}
\end{figure}

In the rest of this section, the \matHRF material is studied and compared to \matS material in terms of variograms. The \matHRF material, with both material parameters $E$ and $\nu$ ($\eta_E=\eta_{\nu}=0.7$) fluctuating according to realizations of two independent random fields, was subjected to \emph{Load 2}. The variogram of $\sigma_{xx}$ produced by the GSTools python package \cite{GSTools} is shown in Fig.~\ref{fig:vario_0} along with the best fits by a~square exponential covariance model. It is seen that both \matHR and \matVRD material variants exhibit similar spatial dependency as the \matS material, but the \matHRF material variant gives significantly lower variances. The \matHRF variograms get closer to the other material variants with decreasing autocorrelation length $
\lcorr$, however, there is a~limit given by the size of the rigid bodies under which the autocorrelation cannot be meaningfully represented. In the attempt to replicate the \matS material, the spatial correlation shall not be considered.

\subsection{Combined deviatoric and volumetric load}

Previously studied separately, the volumetric and deviatoric loads are now combined. Table~\ref{tab:volum_load_vector} shows the macroscopic strain tensor defining \emph{Load 4} as a~sum of volumetric \emph{Load 1} and deviatoric \emph{Load 3}. Since the material is linearly elastic, the results are simply a~sum of the earlier results obtained for \emph{Load 1} and \emph{3}.  Therefore, stress oscillations in the material variants \matS, \matHRD, and \matVRD under \emph{Load 4} are equal to those under \emph{Load 3}, because volumetric loading renders a~homogeneous stress field. In contrast, the \matHR material variant sums up two non-zero fields of oscillation. Once again, it proves that no \matHR randomization can replicate the standard material for an~arbitrary loading vector, since the volumetric component affects the stress oscillation differently.

Due to the different approaches to stress homogenization discussed earlier, the stress oscillations vary between the \matHRD and \matVRD materials, even when subjected to the same randomization of elastic parameters and loads. In the case of \emph{Load 3} and subsequently \emph{Load 4} and the randomization prescribed by $\eta_D = 0.8$, the standard deviation of the component $\sigma_{xx}$ is higher for \matHRD compared to \matVRD ($7.621 \times 10^{4}$ versus $5.758 \times 10^{4}$). This trend holds for all other stress components, although the ratios of these values differ slightly between the stress components, as exemplified by the component $\sigma_{xz}$ ($3.266 \times 10^{4}$ versus $2.905 \times 10^{4}$).

\begin{figure}[tp!]
\centering
\includegraphics[width=6.5in]
{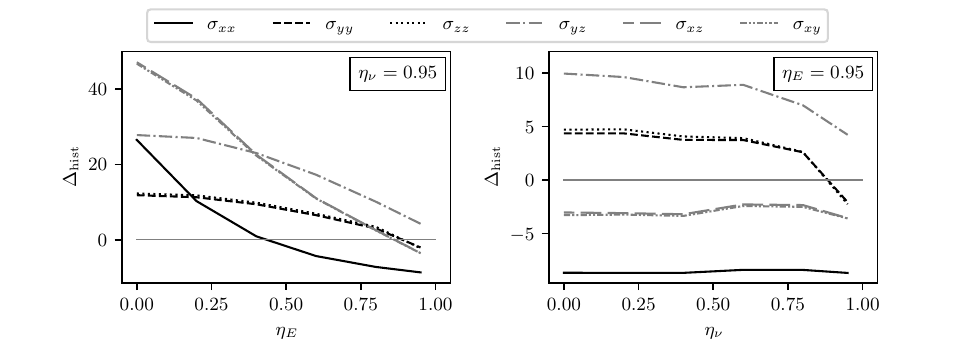}
\caption{Histogram distance $\Delta_{\mathrm{hist}}$ between \matS and \matHR material variants for different $\eta_E$ and $\eta_{\nu}$ parameters of randomization for \emph{Load 5}. On the left at constant $\eta_{\nu} = 0.95$, on the right at constant $\eta_{E} = 0.95$.
}
\label{fig:metric_volumetric}
\end{figure}

\section{Inelastic behavior}
The second part of this research is devoted to the fracture behavior of the preceding models.  For \emph{physical} discretizations of concrete, the lattice configuration corresponds to material structure. A~prime example is the LDPM. Whereas \emph{non-physical} discretization schemes bear no relation to material structure, they possess some desirable properties of continuum formulations (e.g., elastic homogeneity), while retaining the benefits of a~discrete formulation when simulating fracture.

The inelastic constitutive model described in Sec.~\ref{sec:inelastic} corresponds now to the standard material (\matS). The tensile strength $\ft = 2.8$\,MPa and fracture energy $\Gt = 40$\,J/m$^2$ are selected as typical values for structural concrete. The material variant \matH becomes the same inelastic material model extended by the projected eigenstrains according to Eq.~\eqref{eq:eigenstrain}, where the secant modulus $(1-d)E$ is substituted for elastic modulus $E$. In the elastic regime, the elastic \matH material is exactly recovered. The behavior of the inelastic material retains its damage-based character as the unloading occurs linearly toward the origin of the stress-strain space. 

The periodic RVE is subjected to \emph{Load 5}, an~increasing macroscopic strain $\varepsilon_{xx}$ while all shear strains are set to zero $\gamma_{yz} = \gamma_{xz} = \gamma_{xy} = 0$ and transverse normal stresses are also zero, $\sigma_{yy} = \sigma_{zz} = 0$. Figure~\ref{fig:metric_volumetric} demonstrates that arguably the best fit for this loading, in terms of stress oscillation in the elastic regime for the material \matHR, is $\eta_E=\eta_{\nu}=0.95$. This analysis is identical to the one performed in Fig.~\ref{fig:nu_random_metric} except the loading is different. Once again, it confirms that randomization that at least partially resembles the standard material is dictated by the imposed load. Nevertheless, these $\eta$ parameters are used hereinafter for the inelastic version of the \matHR material.

\subsection{\matS and \matHR material variants}
\begin{figure}[tp!]
\centering
\includegraphics[width=3.2in]
{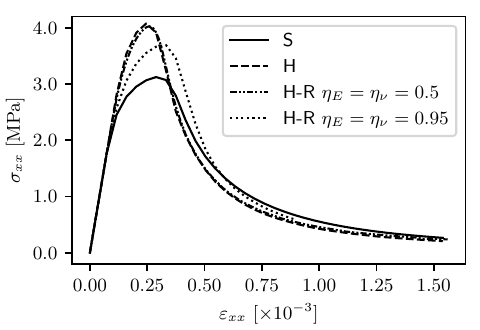}
\includegraphics[width=3.2in]{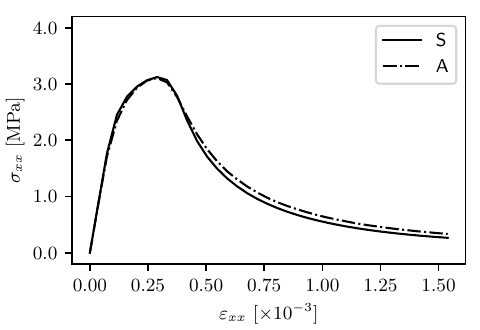}\hfill
\caption{Stress-strain diagram computed as an~average from 100 RVE responses to uniaxial strain loading; left: standard (\matS) material compared to homogeneous (\matH) and randomized (\matHR) variants; right: adjusted (\matA) material matching the standard one.}
\label{fig:50_avg_LD}
\end{figure}

\begin{figure}[tp!]
\centering
\includegraphics[width=3.2in]
{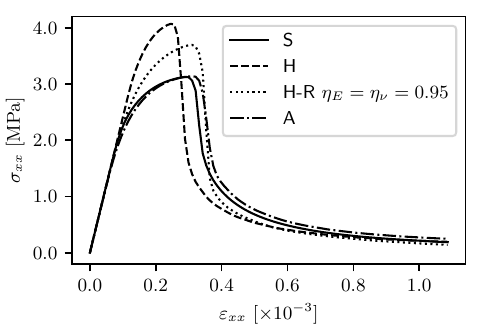}\hfill
\caption{Stress strain-diagram of a~single RVE of size $\lRVE=100$\,mm subjected to \emph{Load 5}.}
\label{fig:100_2_LD}
\end{figure}

Figure \ref{fig:50_avg_LD} shows the average stress $\sigma_{xx}$ evolving with imposed strain $\varepsilon_{xx}$, computed on 100 RVEs of size $\lRVE=50$\,mm. The smaller size was used for the sake of reducing computational time. On the left-hand side the average responses of standard material (\matS), homogeneous material (\matH), and two instances of randomized material (\matHR) are compared. The material \matH remains elastic longer as there is no randomness in the elastic phase. However, after achieving the maximum load, the \matH material becomes more brittle as redistribution mechanisms due to heterogeneity are missing. Adding a~moderate degree of randomness through $\eta_{\nu}=\eta_E=0.5$ does not greatly improve the correspondence, although the earlier commencement of the inelastic phase, a~lower peak load and slightly higher toughness are evident. For $\eta_{\nu}=\eta_E=0.95$ the difference is already substantial, but still far from the standard material. Since there is little margin for widening the uniform distribution (for $\eta>1$ negative values are reached), the fracture parameters of the \matHR material are adjusted manually.  By trial and error, it was found that the average response of the \matHR material with $\eta_{\nu}=\eta_E=0.95$ and adjusted tensile strength to $0.8\ft=2.24$\,MPa and fracture energy in tension to $1.1\Gt=44$\,J/m$^2$ matches closely the average response of the standard material. This optimally {\bf adjusted} material variant will designated as {\tt A} hereafter. The right side of Fig.~\ref{fig:50_avg_LD} shows the matching mean responses of the \matS and \matA materials. Figure~\ref{fig:100_2_LD} shows the stress-strain diagrams computed on larger RVE ($\lRVE=100$\,mm) to verify that the described trends are independent of the RVE size.  The same randomization of the elastic parameters and the same adjustments of the inelastic parameters are applied. Although the shape of the diagrams is size dependent and therefore different from Fig.~\ref{fig:50_avg_LD}, the qualitative differences are preserved and the \matS and \matA materials still display similar responses.

\begin{figure}[tb!]
\centering
\begin{subfigure}{0.32\textwidth}
\includegraphics[trim={1mm 1mm 2cm 3cm},clip,width=1.\textwidth]
{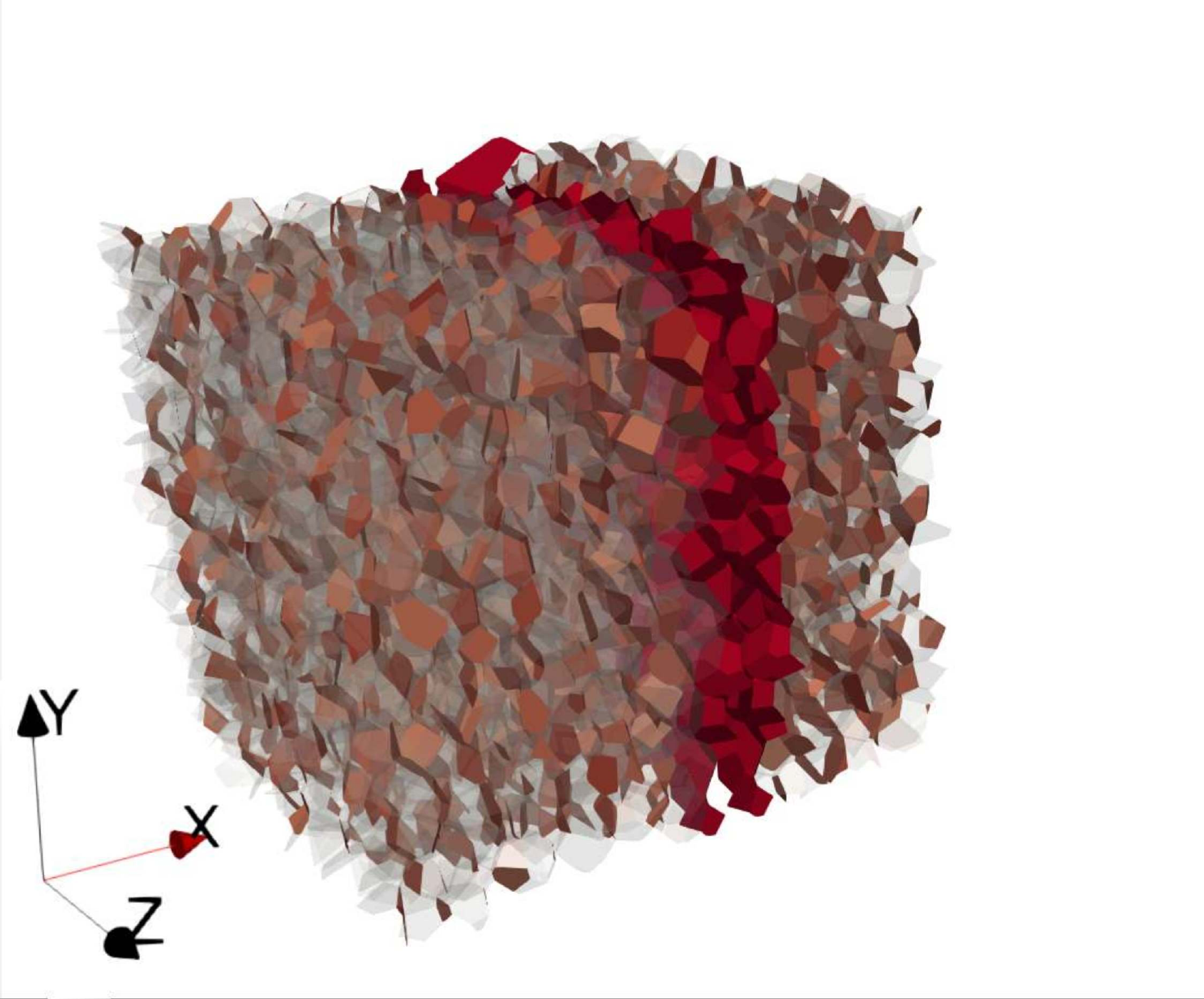}
\caption{\matS material}
\end{subfigure}
\begin{subfigure}{0.32\textwidth}
\includegraphics[trim={1.5cm 0 1.5cm 2cm},clip,width=1.\textwidth]
{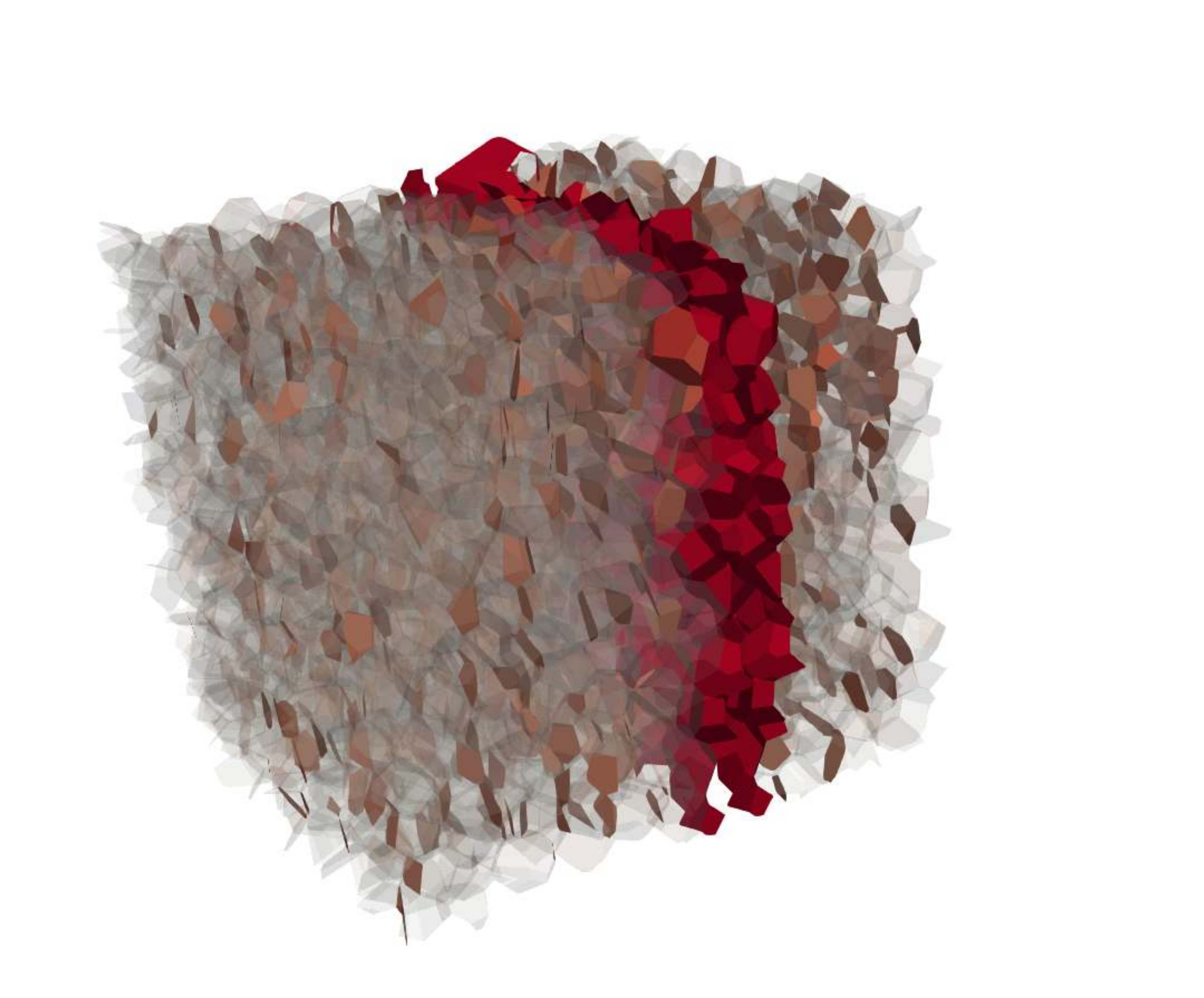}
\caption{\matH material}
\end{subfigure}
\begin{subfigure}{0.32\textwidth}
\includegraphics[trim={3cm 0 1mm 2cm},clip,width=1.\textwidth]
{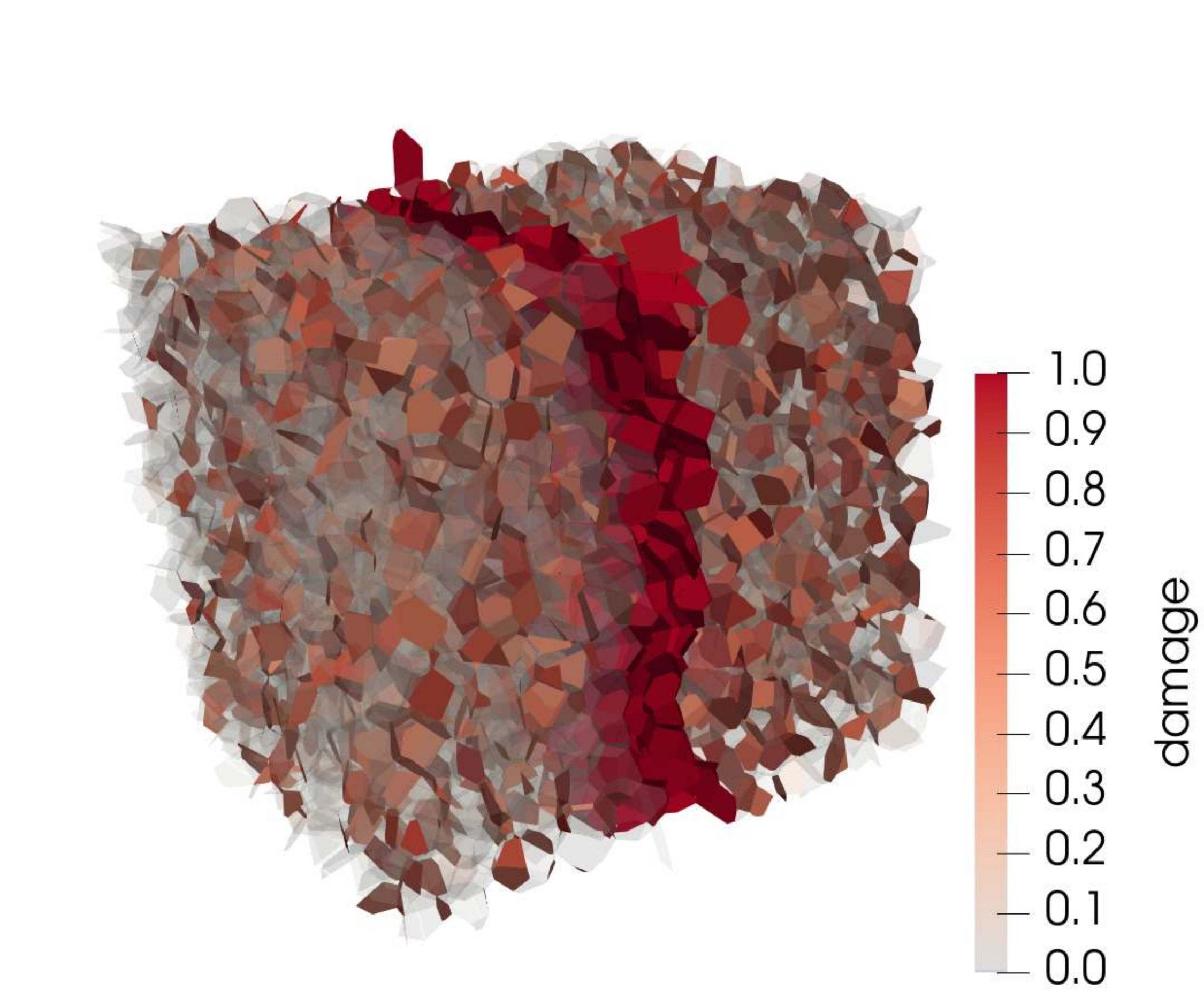}
\caption{\matA material}
\end{subfigure}
\caption{Damage distribution in RVEs at the end of tensile loading from Fig. \ref{fig:100_2_LD}, loading direction is along the $x$ axis.}
\label{fig:paraview_damage}
\end{figure}

Once the adjusted material (\matA) was created, we can study the differences of its fracture process from the fracturing of the \matS material. Figure~\ref{fig:paraview_damage} shows the damage distribution in the RVE of size $\lRVE=100$\,mm at the end of the simulation for the materials \matS, \matH and \matA. All of them exhibit similar damage concentrations in the macroscopic fracture region perpendicular to the loading direction. The material \matH shows less damage outside the crack compared to the materials \matS and \matA. This is in agreement with the stress-strain diagram, which indicates that both \matS and \matA materials undergo larger inelastic development prior to the peak stress at which the localization of strains into a~macro-crack occurs. The dissipated energy is evaluated from the external work using the macroscopic stress and strain tensors. The small amount of the residual elastic strain energy remaining after the terminal time step is neglected. The computed values are 1.29, 1.30, 1.30, and 1.41\,J/m$^3$ for the \matS, \matH, \matHR, and \matA materials, respectively. The higher value observed for \matA compared to \matHR aligns with the difference in the applied mesoscopic parameter $\Gt$. The same analysis, averaged over the set of 50\,mm RVE simulations, yields dissipated energy values 0.26, 0.25, 0.25 and 0.27\,J/m$^3$. It therefore seems that the energy dissipation is not affected by the randomization.

\begin{figure}[tb!]
\centering
\begin{subfigure}{1.5in}
\includegraphics[width=1.\textwidth]
{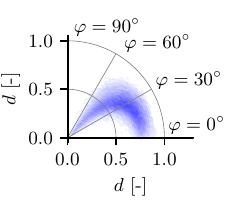}\hfill
\caption{ \matS material, peak load}
\end{subfigure}
\begin{subfigure}{1.4in}
\includegraphics[width=1.\textwidth]{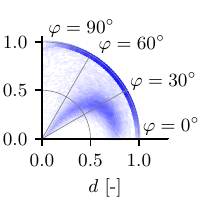}\hfill
\caption{\matS material, final step}
\end{subfigure}
\begin{subfigure}{1.4in}
\includegraphics[width= 1.\textwidth]{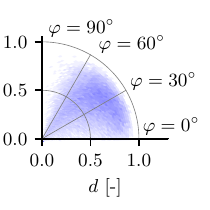}\hfill
\caption{\matA material, peak load}
\end{subfigure}
\begin{subfigure}{1.4in}
\includegraphics[width= 1.\textwidth]{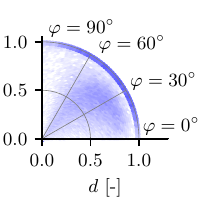}\hfill
\caption{\matA material, final step}
\end{subfigure}
\begin{subfigure}{0.6in}
\includegraphics[width= 1.\textwidth]{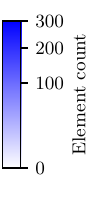}
\caption*{}
\end{subfigure}
\caption{Two dimensional histograms where distance from the center indicates damage variable and angle $\varphi$ is formed by the element normal and the loading direction.}
\label{fig:circle_hist}
\end{figure}

The \matA material shows a~similar amount of distributed damage as the \matS material; the hardening phase of both materials is also similar, as shown in Fig.~\ref{fig:100_2_LD}. However, the character of the distributed damage differs. In the material \matS, the value of damage in an~element is closely related to the orientation of the element with respect to the load direction. On the contrary, Fig.~\ref{fig:circle_hist} demonstrates that the relationship between damage in the \matA material and the orientation of the element is weakened, both at peak load and when approaching a~traction-free crack condition. In the four histograms in Fig.~\ref{fig:circle_hist}, the distance from the center corresponds to the damage value of the element and the angular distance from the horizontal direction is the angle $\varphi$ between the normal direction of the element, $\boldn_N$, and the loading direction along the $x$ axis. This domain is divided into bins, and the depth of the blue color indicates the number of elements in the bins over 30 RVE simulations. At peak load, there are no fully damaged elements and almost no partially damaged elements with $\varphi>70^{\circ}$. The \matS material exhibits a~narrow range of damage dependent on element orientation. The material \matA follows this dependence, but the range is significantly larger. At the end of simulations, as the models approach the traction-free crack condition, one can see a~large number of elements with damage value close to 1. These elements are associated with the macroscopic crack. 

\subsection{\matHRF material variant}
The previous inelastic simulations considered only the spatially independent randomization of elastic parameters. To study the effect of spatial correlation of the elastic parameters on the inelastic response, random fields of several correlation lengths, $\lcorr$, were generated and applied in the \matHRF material variant. 
The maximum stress $\max(\sigma_{xx})$ and the dissipated energy $E_{\mathrm{dis}}$ are computed for $\lcorr$ between 7\,mm and 50\,mm on RVEs of size $\lRVE=100$\,mm. Following from the preceding section, the same adjustments are made to the tensile strength and the fracture energy in tension: $\ft = 2.24$\,MPa and $\Gt = 44$\,J/m$^2$. The probability distribution is still uniform with $\eta_{\nu}=\eta_{E}=0.95$.

\begin{figure}[tb]
\centering
\includegraphics[width=6.5in]
{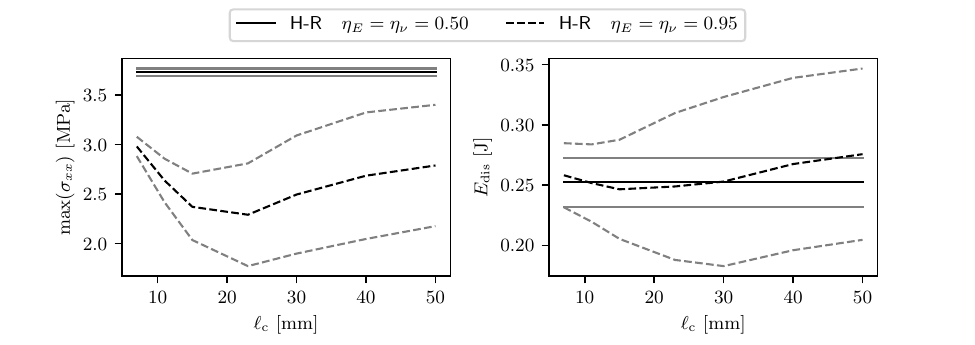}
\caption{Maximum stress $\max(\sigma_{xx})$ and dissipated energy $E_{\mathrm{dis}}$, mean and band of $\pm$ one standard deviation for varying $\lcorr$.}
\label{fig:px_gt}
\end{figure}

Figure~\ref{fig:px_gt} shows the mean values and bands of $\pm$ one standard deviation of $\max(\sigma_{xx})$ and $E_{\mathrm{dis}}$ based on 100 simulations. The results closely resemble the results obtained and interpreted in Refs.~\parencite{EliVor20,VorEli20b}, in which the randomized variables were local strength and fracture energy, and in Ref.~\parencite{SykTej-15} where material strength was considered random. There is a~minimum value of macroscopic strength appearing at a~critical correlation length. This length is long enough so that the stress fluctuations are not \emph{averaged} out within the fracture process zone, but also 
short enough so the \emph{weakest link} effect can take place. For longer correlation lengths, the weakest link model is gradually suppressed, so the strength increases. Asymptotically, material behavior approaches behavior of a~random but spatially constant material.  The variability of the strength increases with $\lcorr$ as well because the averaging within the fracture process zone is reduced and the weakest link effect diminishes. For shorter $\lcorr$ the averaging within the fracture process zone reduces the apparent material variability, and therefore the strength increases while its variability diminishes. In an~extreme case of $\lcorr\rightarrow 0$ the response of the material \matHRF should approach the \matHR response, but discretization into finite-sized rigid bodies prevents the usage of random fields with $\lcorr$ below 7\,mm.

Similar effects can also be seen for the dissipated energy. There is a~critical correlation length for which the dissipation is minimized. In addition, the standard deviation increases with $\lcorr$. Both of these effects are milder than what was seen in strength. Since the tensile fracture energy at the contacts is constant, the observed behavior must be attributed to a~different fracture character. At critical $\lcorr$, the macroscopic crack is believed to be the most localized, so the energy dissipation is the lowest. 

\section{Conclusions}
 Stress oscillations in discrete models due to material heterogeneity are created via two different strategies. In the \emph{standard} approach, they arise from geometric structure of the model. In the \emph{homogeneous} strategy they are governed
by (i) homogenizing the local response of the discrete structure and subsequently (ii) introducing a~controlled degree of randomness through assignment of elastic material parameters in space. Such stress oscillations are of special interest as they influence (and in some situations govern) fracture behavior. Several randomization techniques are used to introduce heterogeneity. Stress analyses are conducted using representative volume elements of the material structure. In particular, the stress fields within homogeneous models, supplemented by heterogeneity, are compared with those based on physical discretizations of the material structure and standard constitutive relations.

The following conclusions are drawn based on a~large number of numerical simulations performed on discrete periodic structures. For the elastic regime, the particular conclusions reads:
\begin{itemize}

\item Two means of homogenizing local behavior (that is, the volumetric-deviatoric decomposition of the constitutive model and the auxiliary stress projection method) are implemented,  producing elastically homogeneous materials. It is demonstrated that they are essentially identical.
\item For a~pure \emph{volumetric} loading both the standard and homogeneous materials produce homogeneous stress fields. As expected, spatial randomization of the macroscopic elastic modulus and Poisson's ratio parameters leads to inhomogeneous stress fields. However, randomizing only the deviatoric part of the constitutive model preserves the homogeneous stress state. 

\item  For a~specific loading configuration (combining both volumetric and deviatoric components), randomization of the elastic parameters of the homogeneous material may accurately correspond to the standard material. However, there appears to be no randomization method that recovers the statistical characteristics of the stress field exhibited by the standard material model across various loading scenarios.

\item Moreover, randomization that produces reasonably similar stress fields may not exist for a~particular loading scenario. Firstly, it might be impossible to match the distributions of all the tensorial stress components for the standard and randomized materials. Secondly, the standard material model generates statistically independent stress components while the randomized material exhibits significant correlations between them. Consequently, rotation of the coordinate system changes the stress components in these materials differently.

\item The spatial correlation of the stress field produced by the standard material is best reproduced by randomizing the elastic parameters of each discrete contact independently. When using spatial dependence in the form of a~random field with correlation length larger or equal to the size of rigid bodies, the resulting correlation lengths of the stress tensor components are substantially larger compared to those of the standard material.
\end{itemize}

Several additional conclusions can be drawn regarding the fracture behavior of these models:
\begin{itemize}
\item Because the homogeneous material effectively eliminates all heterogeneity in the elastic regime, it enters the inelastic domain at strains larger than those of the standard material. Furthermore, its strength is larger and the softening phase is steeper. 

\item Randomization of elastic material properties of the homogeneous material introduces heterogeneity and therefore leads to an~earlier start of the inelastic effects, a~lower strength, and a~milder slope of the softening phase. Nevertheless, even a~large variability in elastic parameters does not introduce sufficient heterogeneity to reproduce the macroscopic behavior of the standard material. 

\item The fracture parameters of the homogeneous material with randomized elasticity can be adjusted so that its inelastic response matches the response of the standard material. Even then, there are significant differences in terms of orientations of the damaged contacts before the macroscopic crack appears. Damage in standard material is closely associated with contact orientation, whereas homogeneous material  exhibits this relationship to a~lesser degree.

\item When the elastic parameters are randomized with spatial correlation, there is a~critical correlation length for which the strength is minimal. This pattern corresponds to a~behavior previously observed~\parencite{EliVor-20}, explained by a~combination of the weakest link effect and the averaging of stress within the fracture process zone. This time, a~milder but essentially identical effect is also seen for energy dissipation, which is also minimized around the critical correlation length. 
\end{itemize}
In short, the two methods of introducing heterogeneity into the model result in qualitatively different local behaviors. Adequate experimental procedures and results are necessary to provide insight into the appropriateness of the two strategies. The implications of this research extend beyond the discrete models domain, as randomization of homogeneous models to fabricate heterogeneous internal structure is used across all modeling approaches.

\section*{Acknowledgement}

Jan Eliáš and Jan Raisinger acknowledge financial support provided by the Czech Ministry of Education, Youth and Sports under project No.~LUAUS24260 and by project INODIN (Innovative methods of materials diagnostics and monitoring of engineering infrastructure to increase its durability and service time -- CZ.02.01.01/00/23\_020/0008487) co-funded by European Union. The former project financed the development of the homogenization technique and implementation of the RVE model, the latter was used to implement other material models and conduct the statistical study. For the purpose of Open Access, a~CC BY 4.0 public copyright license has been applied by the authors to the present document and will be applied to all subsequent versions up to the Author Accepted Manuscript arising from this submission.

\section*{Data availability}

Computational models, scripts to run them, and their results are available at Zenodo repository under DOI \href{https://doi.org/10.5281/zenodo.15303637}{10.5281/zenodo.15303637}.

\printbibliography

@book{Kar49,
	author = {Karhunen, K.},
	series = {Annales Academiae scientiarum Fennicae},
	title = {Zur Spektraltheorie stochastischer Prozesse},
	year = {1946}}

@article{SykTej-15,
title = {{FE} investigations of the effect of fluctuating local tensile strength on coupled energetic–statistical size effect in concrete beams},
journal = {Engineering Structures},
volume = {103},
pages = {239--259},
year = {2015},
issn = {0141-0296},
doi = {10.1016/j.engstruct.2015.09.011},
author = {E. Syroka-Korol and J. Tejchman and Z. Mróz},
}

@article{BhaGom-21,
author = {T. Bhaduri  and S. Gomaa  and M. Alnaggar },
title = {Coupled Experimental and Computational Investigation of the Interplay between Discrete and Continuous Reinforcement in Ultrahigh Performance Concrete Beams. II: Mesoscale Modeling},
journal = {Journal of Engineering Mechanics},
volume = {147},
number = {9},
pages = {04021050},
year = {2021},
doi = {10.1061/(ASCE)EM.1943-7889.0001941},
}

@article{SchGar97,
title = {Fracture simulations of concrete using lattice models: Computational aspects},
journal = {Engineering Fracture Mechanics},
volume = {57},
number = {2},
pages = {319-332},
year = {1997},
issn = {0013-7944},
doi = {10.1016/S0013-7944(97)00010-6},
author = {E. Schlangen and E.J. Garboczi},
}

@article{ZhoAyd-24,
title = {A lattice modelling framework for fracture-induced acoustic emission wave propagation in concrete},
journal = {Engineering Fracture Mechanics},
volume = {312},
pages = {110589},
year = {2024},
issn = {0013-7944},
doi = {10.1016/j.engfracmech.2024.110589},
author = {Y. Zhou and B. B. Aydin and F. Zhang and M.A.N. Hendriks and Y. Yang},
}

@article{BazTab-90,
author = {Z. P. Bažant  and M. R. Tabbara  and M. T. Kazemi  and G. Pijaudier‐Cabot },
title = {Random Particle Model for Fracture of Aggregate or Fiber Composites},
journal = {Journal of Engineering Mechanics},
volume = {116},
number = {8},
pages = {1686--1705},
year = {1990},
doi = {10.1061/(ASCE)0733-9399(1990)116:8(1686)},
}

@article{KikKaw-92,
title = {The rigid bodies—spring models and their applications to three-dimensional crack problems},
journal = {Computers \& Structures},
volume = {44},
number = {1},
pages = {469-480},
year = {1992},
issn = {0045-7949},
doi = {10.1016/0045-7949(92)90269-6},
author = {A. Kikuchi and T. Kawai and N. Suzuki},
}

@article{DonBri-24,
author = {Jia, D. and Brigham, J. C. and Fascetti, A.},
title = {An efficient static solver for the Lattice Discrete Particle Model},
journal = {Computer-Aided Civil and Infrastructure Engineering},
volume = {39},
number = {23},
pages = {3531-3551},
doi = {10.1111/mice.13306},
year = {2024}
}

@article{Gra23,
title = {{3D} lattice meso-scale modelling of the effect of lateral compression on tensile fracture processes in concrete},
journal = {International Journal of Solids and Structures},
volume = {262-263},
pages = {112086},
year = {2023},
issn = {0020-7683},
doi = {10.1016/j.ijsolstr.2022.112086},
author = {P. Grassl},
}

@ARTICLE{ZhuPat-22,
	author = {Zhu, Z. and Pathirage, M. and Wang, W. and Troemner, M. and Cusatis, Gianluca},
	title = {Lattice discrete particle modeling of concrete under cyclic tension–compression with multi-axial confinement},
	journal = {Construction and Building Materials},
volume = {352},
pages = {128985},
year = {2022},
issn = {0950-0618},
	doi = {10.1016/j.conbuildmat.2022.128985},
}

@article{ChaCif-17,
title = {Fracture analyses using spring networks with random geometry},
journal = {Materials},
volume = {10},
number = {2},
pages = {207},
year = {2017},
issn = {1996-1944},
doi = {10.3390/ma10020207},
author = {Montero-Chacón, Francisco and Héctor Cifuentes and Fernando Medina},
}

@article{BolSai98,
title = {Fracture analyses using spring networks with random geometry},
journal = {Engineering Fracture Mechanics},
volume = {61},
number = {5},
pages = {569-591},
year = {1998},
issn = {0013-7944},
doi = {10.1016/S0013-7944(98)00069-1},
author = {J.E. Bolander and S. Saito},
}

@article{SchMie92,
title = {Simple lattice model for numerical simulation of fracture of concrete materials and structures},
journal = {Materials and Structures},
volume = {25},
pages = {534--542},
year = {1992},
issn = {1359--5997},
doi = {10.1007/BF02472449},
author = {Schlangen, E. and van Mier, J.G.M.},
}

@article{YinTro-24,
title = {An interprocess communication-based two-way coupling approach for implicit–explicit multiphysics lattice discrete particle model simulations},
journal = {Engineering Fracture Mechanics},
volume = {310},
pages = {110515},
year = {2024},
issn = {0013-7944},
doi = {10.1016/j.engfracmech.2024.110515},
author = {H. Yin and M. Troemner and W. Li and E. Lale and L. Yang and L. Shen and M. Alnaggar and G. {Di Luzio} and G. Cusatis},
}

@article{MiuNak-23,
title = {Expansive spalling mechanism of concrete due to high temperature based on developed hygro-thermal-mechanical model by {3D-RBSM-TNM}},
journal = {Engineering Fracture Mechanics},
volume = {284},
pages = {109216},
year = {2023},
issn = {0013-7944},
doi = {10.1016/j.engfracmech.2023.109216},
author = {Taito Miura and Hikaru Nakamura and Yoshihito Yamamoto},
}

@article{LuoAsa-23,
title = {Mesoscale simulation of compression-induced cracking and failure of {ASR}-damaged concrete with stirrup confinement},
journal = {Engineering Fracture Mechanics},
volume = {277},
pages = {108977},
year = {2023},
issn = {0013-7944},
doi = {10.1016/j.engfracmech.2022.108977},
author = {J. Luo and S. Asamoto and K. Nagai},
}

@article{Spa89,
	author = {Spanos, P. D. and Ghanem, R. G.},
	journal = {J. Eng. Mech., ASCE},
	number = {5},
	pages = {1035--1053},
	publisher = {American Society of Civil Engineers},
	title = {Stochastic finite element expansion for random media},
	volume = {115},
	year = {1989}}

@article{Ste09,
	author = {G. Stefanou},
	journal = {Comp. Meth. Appl. Mech. Eng.},
	number = {2009},
	pages = {1031-1051},
	title = {{The stochastic finite element method: Past, present and future}},
	volume = {198},
	year = {2009}}

@book{Gha03,
	author = {Ghanem, R. G. and Spanos, P. D.},
	publisher = {Courier Corporation},
	title = {Stochastic finite elements: a spectral approach},
	year = {2003}}

@article{LiLu-08,
	title = {Nataf transformation based point estimate method},
	volume = {53},
	number = {17},
	journal = {Chin Sci Bull},
	author = {Li, {H.} and Lü, {Z.} and Yuan, {X.}},
	year = {2008},
	pages = {2586--2592},
    doi = "10.1007/s11434-008-0351-0"
}

@Article{Vor08,
  author =   {M. Vo{\v{r}}echovsk{\'{y}}},
  title =    {Simulation of simply cross correlated random fields by series expansion methods},
  journal =  {Struct Saf},
  year =     {2008},
  pages =    {337--363},
  doi    =   {10.1016/j.strusafe.2007.05.002},
  volume =   {30},
  number =   {4},
  issn =     {0167-4730},
}

@article{VorNov09,
	title = {Correlation control in small-sample Monte Carlo type simulations {I}: A simulated annealing approach},
	volume = {24},
	number = {3},
	journal = {Probabilist Eng Mech},
	author = {Vo\v{r}echovsk\'{y}, M. and Nov\'{a}k, D.},
	year = {2009},
	pages = {452--462},
        issn = "0266-8920",
        doi = "10.1016/j.probengmech.2009.01.004",
}

@article{NicHad-13,
title = {On the definition of the stress tensor in granular media},
journal = {International Journal of Solids and Structures},
volume = {50},
number = {14},
pages = {2508--2517},
year = {2013},
issn = {0020-7683},
doi = {10.1016/j.ijsolstr.2013.04.001},
author = {F. Nicot and N. Hadda and M. Guessasma and J. Fortin and O. Millet},
}

@article{YanReg19,
title = {Definition and symmetry of averaged stress tensor in granular media and its {3D} {DEM} inspection under static and dynamic conditions},
journal = {International Journal of Solids and Structures},
volume = {161},
pages = {243--266},
year = {2019},
issn = {0020-7683},
doi = {10.1016/j.ijsolstr.2018.11.021},
author = {B. Yan and R. A. Regueiro},
}

@article{LiKiu93,
	title = {Optimal Discretization of Random Fields},
	volume = {119},
	number = {6},
	author = {Li, Ch. and Der Kiureghian, A.},
	year = {1993},
	pages = {1136--1154},
	journal = {J Eng Mech - ASCE},
        issn={0733-9399},
        doi={10.1061/(ASCE)0733-9399(1993)119:6(1136)}
}

@article{DreDeJ72,
title = {Photoelastic verification of a mechanical model for the flow of a granular material},
journal = {Journal of the Mechanics and Physics of Solids},
volume = {20},
number = {5},
pages = {337--340},
year = {1972},
issn = {0022-5096},
doi = {10.1016/0022-5096(72)90029-4},
author = {A. Drescher and G. {de Josselin de Jong}},
}

@article{CusPel-11,
	title = {Lattice Discrete Particle Model ({LDPM}) for failure behavior of concrete. {I}: Theory},
	journal = {Cement and Concrete Composites},
	volume = {33},
	number = {9},
	pages = {881--890},
	year = {2011},
	issn = {0958-9465},
	doi = {10.1016/j.cemconcomp.2011.02.011},
	author = {G. Cusatis and D. Pelessone and A. Mencarelli},
}

@article{BolEli-21,
author = {J. E. Bolander and J. Eli\'{a}\v{s} and G. Cusatis and K. Nagai},
title = {Discrete mechanical models of concrete fracture},
journal = {Engineering Fracture Mechanics},
year = {2021},
volume = {257},
pages = {108030},
doi = {10.1016/j.engfracmech.2021.108030},
issn = {0013-7944}
}

@article{EliVor20,
author = {J. Eli\'{a}\v{s} and M. Vo\v{r}echovsk\'{y}},
title = {Fracture in random quasibrittle media: {I}. {D}iscrete mesoscale simulations of load capacity and fracture process zone},
journal = {Engineering Fracture Mechanics},
year = {2020},
volume = {235},
pages = {107160},
doi = {10.1016/j.engfracmech.2020.107160},
issn = {0013-7944}
}

@article{VorEli20b,
author = {M. Vo\v{r}echovsk\'{y} and J. Eli\'{a}\v{s}},
title = {Fracture in random quasibrittle media: {II}. {A}nalytical model based on extremes of the averaging process},
journal = {Engineering Fracture Mechanics},
year = {2020},
volume = {235},
pages = {107155},
doi = {10.1016/j.engfracmech.2020.107155},
issn = {0013-7944}
}

@article{Eli16,
author = {J. Eli\'{a}\v{s}},
title = {Adaptive technique for discrete models of fracture},
journal = {International Journal of Solids and Structures},
year = {2016},
volume = {100--101},
pages = {376--387},
doi = {10.1016/j.ijsolstr.2016.09.008},
issn = {0020-7683}
}

@article{EliVor-15,
author = {J. Eli\'{a}\v{s} and M. Vo\v{r}echovsk\'{y} and J. Sko\v{c}ek and Z. P. Ba\v{z}ant},
title = {Stochastic discrete meso-scale simulations of concrete fracture: comparison to experimental data},
journal = {Engineering Fracture Mechanics},
year = {2015},
volume = {135},
pages = {1--16},
doi = {10.1016/j.engfracmech.2015.01.004},
issn = {0013-7944}
}

@article{YanTro-24,
  title = {A novel analytical model of particle size distributions in granular materials},
  author = {Yang, L. and Troemner, M. and Cusatis, G. and Su, H.},
  year  = 2024,
  journal = {Engineering with Computers},
  doi = {10.1007/s00366-024-02042-7},
  issn = {1435-5663},
}

@article{VorEli20,
author = {M. Vo\v{r}echovsk\'{y} and J. Eli\'{a}\v{s}},
title = {Modification of the Maximin and $\phi_p$ (phi) criteria to achieve statistically uniform distribution of sampling points},
journal = {Technometrics},
year = {2020},
volume = {62},
number = {3},
pages = {371--386},
doi = {10.1080/00401706.2019.1639550},
issn = {0040-1706}
}

@article{EliVor-20,
author = {J. Eli\'{a}\v{s} and M. Vo\v{r}echovsk\'{y} and V. Sad\'{i}lek},
title = {Periodic version of the minimax distance criterion for Monte Carlo integration},
journal = {Advances in Engineering Software},
year = {2020},
volume = {149},
pages = {102900},
doi = {10.1016/j.advengsoft.2020.102900},
issn = {0965-9978}
}

@article{RezZho-17,
	title = {Adaptive multiscale homogenization of the lattice discrete particle model for the analysis of damage and fracture in concrete},
	journal = {International Journal of Solids and Structures},
	volume = {125},
	pages = {50-67},
	year = {2017},
	issn = {0020-7683},
	doi = {h10.1016/j.ijsolstr.2017.07.016},
	author = {R. Rezakhani and X. Zhou and G. Cusatis}
}

@article{RezCus16,
	title = "Asymptotic expansion homogenization of discrete fine-scale models with rotational degrees of freedom for the simulation of quasi-brittle materials",
	journal = "Journal of the Mechanics and Physics of Solids",
	volume = "88",
	pages = "320--345",
	year = "2016",
	issn = "0022-5096",
	doi = "10.1016/j.jmps.2016.01.001",
	author = "R. Rezakhani and G. Cusatis"
}

@article{EliCus22,
author = {J. Eli\'{a}\v{s} and G. Cusatis},
title = {Homogenization of discrete mesoscale model of concrete for coupled mass transport and mechanics by asymptotic expansion},
journal = {Journal of the Mechanics and Physics of Solids},
year = {2022},
volume = {167},
pages = {105010},
doi = {10.1016/j.jmps.2022.105010},
issn = {0022-5096}
}

@article{EliVor16,
author = {J. Eli\'{a}\v{s} and M. Vo\v{r}echovsk\'{y}},
title = {Modification of the Audze–Eglajs criterion to achieve a uniform distribution of sampling points},
journal = {Advances in Engineering Software},
year = {2016},
volume = {100},
pages = {82--96},
doi = {10.1016/j.advengsoft.2016.07.004},
issn = {0965-9978}
}

@article{MasKve-23,
author = {J. Ma\v{s}ek and J. Kv\v{e}to\v{n} and J. Eli\'{a}\v{s}},
title = {Adaptive discretization refinement for discrete models of coupled mechanics and mass transport in concrete},
journal = {Construction and Building Materials},
year = {2023},
volume = {395},
pages = {132243},
doi = {10.1016/j.conbuildmat.2023.132243},
issn = {0950-0618}
}

@article{CusCed07,
	title = {Two-scale study of concrete fracturing behavior},
	journal = {Engineering Fracture Mechanics},
	volume = {74},
	number = {1},
	pages = {3--17},
	year = {2007},
	note = {Fracture of Concrete Materials and Structures},
	issn = {0013-7944},
	doi = {10.1016/j.engfracmech.2006.01.021},
	author = {G. Cusatis and L. Cedolin},
}

@article{AlnPel-19,
	author = {M. Alnaggar  and D. Pelessone  and G. Cusatis },
	title = {Lattice Discrete Particle Modeling of Reinforced Concrete Flexural Behavior},
	journal = {Journal of Structural Engineering},
	volume = {145},
	number = {1},
	pages = {04018231},
	year = {2019},
	doi = {10.1061/(ASCE)ST.1943-541X.0002230}
}

@article{KryRot96,
	author = {Kruyt, N. P. and Rothenburg, L.},
	title = "{Micromechanical Definition of the Strain Tensor for Granular Materials}",
	journal = {Journal of Applied Mechanics},
	volume = {63},
	number = {3},
	pages = {706--711},
	year = {1996},
	month = {09},
	issn = {0021-8936},
	doi = {10.1115/1.2823353},
}

@article{Web66,
	author = {Weber, J.},
	title = "{Recherches concernant les contraintes intergranulaires dans les
	milieux pulvérulents}",
	journal = {Bulletin de Liaison des Ponts-et-Chaussées},
	volume = {20},
	pages = {1--20},
	year = {1966},
}

@article{Bag96,
	title = {Stress and strain in granular assemblies},
	journal = {Mechanics of Materials},
	volume = {22},
	number = {3},
	pages = {165--177},
	year = {1996},
	issn = {0167-6636},
	doi = {10.1016/0167-6636(95)00044-5},
	author = {K. Bagi},
}

@book{Love1927,
	title     = "A treatise on the mathematical theory of elasticity",
	author    = "A. E. H. Love",
	year      = 1927,
	publisher = "Cambridge University Press",
	volume = "1",	
}

@Article{KuhDadd-00,
author="Kuhl, E. and D'Addetta, G. A. and Herrmann, H. J. and Ramm, E.",
title="A comparison of discrete granular material models with continuous microplane formulations",
journal="Granular Matter",
year="2000",
volume="2",
number="3",
pages="113--121",
issn="1434-5021",
doi="10.1007/s100350050003",
}

@article{Eli20,
author = {J. Eli\'{a}\v{s}},
title = {Elastic properties of isotropic discrete systems: Connections between geometric structure and Poisson’s ratio},
journal = {International Journal of Solids and Structures},
year = {2020},
volume = {191-192},
pages = {254--263},
doi = {10.1016/j.ijsolstr.2019.12.012},
issn = {0020-7683}
}

@article{BatRot88,
title = "Micromechanical Aspects of Isotropic Granular Assemblies With Linear Contact Interactions",
journal = "Journal of Applied Mechanics",
volume = "55",
number = "1",
pages = "17--23",
year = "1988",
issn = "0021-8936",
doi = "10.1115/1.3173626",
author = "Bathurst R.,J. and Rothenburg L."
}

@article{Eli17,
author = {J. Eli\'{a}\v{s}},
title = {Boundary Layer Effect on Behavior of Discrete Models},
journal = {Materials},
year = {2017},
volume = {10},
pages = {157},
doi = {10.3390/ma10020157},
issn = {1996-1944}
}

@article{ZhaEli-24,
author = {Q. Zhang and J. Eli\'{a}\v{s} and K. Nagai and J. E. Bolander},
title = {Discrete Modeling of Elastic Heterogeneous Media},
journal = {Mechanics Research Communications},
volume = {137},
pages = {104277},
year = {2024},
doi = {10.1016/j.mechrescom.2024.104277},
issn = {0093-6413}
}

@article{CusRez-17,
title = {Discontinuous {C}ell {M}ethod ({DCM}) for the simulation of cohesive fracture and fragmentation of continuous media},
journal = {Engineering Fracture Mechanics},
volume = {170},
pages = {1--22},
year = {2017},
issn = {0013-7944},
doi = {10.1016/j.engfracmech.2016.11.026},
author = {G. Cusatis and R. Rezakhani and E. A. Schauffert},
}

@article{AsaKaz-17,
author = {Asahina, D. and Aoyagi, K. and Kim, K. and Birkholzer, J. and Bolander, J.},
year = {2017},
month = {01},
pages = {195-206},
title = {Elastically-homogeneous lattice models of damage in geomaterials},
volume = {81},
journal = {Computers and Geotechnics},
doi = {10.1016/j.compgeo.2016.08.015}
}

@article{AsaIto-15,
	Author = {D. Asahina and K. Ito and J.E. Houseworth and J. T. Birkholzer and J. E. Bolander},
	Doi = {10.1016/j.compgeo.2015.07.013},
	Issn = {0266-352X},
	Journal = {Computers and Geotechnics},
	Pages = {60 - 67},
	Title = {Simulating the Poisson effect in lattice models of elastic continua},
	Volume = {70},
	Year = {2015},
}

@Article{GSTools,
AUTHOR = {M\"uller, S. and Sch\"uler, L. and Zech, A. and He{\ss}e, F.},
TITLE = {\texttt{GSTools} v1.3: a toolbox for geostatistical modelling in Python},
JOURNAL = {Geoscientific Model Development},
VOLUME = {15},
YEAR = {2022},
NUMBER = {7},
PAGES = {3161--3182},
DOI = {10.5194/gmd-15-3161-2022}
}

@book{OkaBoo-00,
author = {Okabe, A. and Boots, B. and Sugihara, K. and Chiu, S.},
year = {2000},
month = {01},
pages = {},
title = {Spatial Tessellations: Concepts and Applications of Voronoi Diagrams},
volume = {43},
isbn = {9780471986355},
journal = {Wiley Series in Probability and Mathematical Statistics, Chichester, New York: Wiley, 1992},
doi = {10.2307/2687299}
}

@article{NagSat2005,
author = {Nagai, K. and Sato, Y. and Ueda, T.},
year = {2005},
pages = {385--402},
title = {Mesoscopic simulation of failure of mortar and concrete by {3D} {RBSM}},
volume = {3},
number = {3},
journal = {Journal of Advanced Concrete Technology},
}

@article{KanBol17,
author = {Kang, J. and Bolander, J. E.},
year = {2017},
pages = {245--261},
title = {Event-based lattice modeling of strain-hardening cementitious composites},
volume = {206},
journal = {International Journal of Fracture},
}

@article{EliCus25,
author = {J. Eli\'{a}\v{s} and G. Cusatis},
title = {Macroscopic stress, couple stress and flux tensors derived through energetic equivalence from microscopic continuous and discrete heterogeneous finite representative volumes},
journal = {Computer Methods in Applied Mechanics and Engineering},
volume = {436},
pages = {117688},
year = {2025},
doi = {10.1016/j.cma.2024.117688},
issn = {0045-7825}
}

@article{Breysse90,
author = {D. Breysse},
title = {Probabilistic formulation of damage-evolution law of cementitious composites},
journal = {Journal of Engineering Mechanics, ASCE},
volume = {116},
number = {7},
pages = {1489--1510},
year = {1990}
}

\end{document}